\documentclass[preprint,12pt]{elsarticle}
\usepackage{amsmath,bm,amssymb,subcaption,graphicx,dsfont,array,hyperref,cleveref,comment}
\usepackage{enumerate}
\usepackage{xcolor}
\usepackage{multirow,rotating}
\usepackage{natbib}
\usepackage{algorithm,algpseudocode} % Algorithms
\usepackage[page]{appendix}
\usepackage{titlesec}
\crefname{appendix}{}{}

\hypersetup{breaklinks=true,
            pdfauthor={},
            pdfkeywords = {},
            pdftitle={},
            colorlinks=true,
            citecolor=blue,
            urlcolor=blue,
            linkcolor=magenta,
            pdfborder={0 0 0}}
            
\newcommand\BibTeX{{\rmfamily B\kern-.05em \textsc{i\kern-.025em b}\kern-.08em
\kern-.1667em\lower.7ex\hbox{E}\kern-.125emX}}



\begin{document}

\begin{titlepage}
    \centering
    \vspace*{1cm}

    \Huge
    \textbf{Fast Emulation, Modular Calibration, and Active Learning for Simulators with Functional Response}
    
    \vspace{1.5cm}
    \Large
    Grant Hutchings\textsuperscript{a,b}\footnote{Corresponding author: Email: grant.hutchings@lanl.gov, Phone: +1 505-667-2200, Present address: P.O. Box 1663 Los Alamos, NM 87545.},
    Derek Bingham\textsuperscript{b},
    Kellin Rumsey\textsuperscript{a},
    Earl Lawrence\textsuperscript{a}

    \vfill

    \textsuperscript{a}Statistical Sciences, Los Alamos National Laboratory, NM, United States\\
    \textsuperscript{b}Department of Statistics, Simon Fraser University, BC, Canada\\

    \vspace{1.5cm}

    \large
    \today

\end{titlepage}

\title{Fast Emulation, Modular Calibration, and Active Learning for Simulators with Functional Response}

\author[cor1,1,2]{Grant Hutchings}
\author[2]{Derek Bingham}
\author[1]{Kellin Rumsey}
\author[1]{Earl Lawrence}

%% Author affiliation

\address[1]{Statistical Sciences, Los Alamos National Laboratory, NM, United States}
\address[2]{Department of Statistics, Simon Fraser University, BC, Canada}

\cortext[cor1]{Corresponding author: Email: grant.hutchings@lanl.gov (Grant Hutchings), Phone: +1 505-667-2200, Present address: P.O. Box 1663 Los Alamos, NM 87545.}

\begin{abstract}

\noindent Scalable surrogate models enable efficient emulation of computer models (or simulators), particularly when dealing with large ensembles of runs. While Gaussian process (GP) models are commonly employed for emulation, they face limitations in scaling to large datasets. Furthermore, when dealing with dense functional output, such as spatial or time-series data, additional complexities arise, requiring careful handling to ensure fast emulation. This work presents a highly scalable emulator for functional data incorporating local Gaussian process regression. The emulator utilizes global GP lengthscale parameter estimates to scale the input space, leading to a substantial improvement in prediction speed. We demonstrate that our fast approximation-based emulator can serve as a viable alternative to a fully Bayesian approach for functional response, while drastically reducing computational costs. The proposed emulator is applied to quickly calibrate a multiphysics continuum hydrodynamics simulator with a large ensemble of 20000 runs. The methods presented are implemented in the \texttt{R} package \href{https://github.com/granthutchings/FlaGP/tree/main}{\texttt{FlaGP}}.

\end{abstract}

%%Research highlights
\begin{highlights}
\item A new approach for fast emulation and calibration of computer models with functional response is developed. The approach is motivated by a need for near-real-time prediction and Bayesian calibration with very large ensemble datasets.

\item An active learning approach for functional response emulators is developed, and it is shown how the proposed emulation framework facilitates rapid updating of integrated prediction variance when new simulation inputs are added.

\end{highlights}


%\def\spacingset#1{\renewcommand{\baselinestretch}%
%{#1}\small\normalsize} \spacingset{1.5}
%\setlength{\abovedisplayskip}{4pt}
%\setlength{\belowdisplayskip}{4pt}
%\vspace{-8mm}

\maketitle

\section{Introduction}\label{sec:intro}

Computer models (or simulators) have become fundamental in the study of complex physical systems. In some cases they are so computationally intensive that only a small number of runs are performed \citep{Sacks89}, while in others they are fast enough that large ensembles of trails are obtained \citep[e.g.,][]{Francom2019InferringAR,compact_cov}. Statistical surrogates, or emulators, remain a key tool enabling efficient exploration of simulator response \citep{gramacy2020surrogates}. In this work our principal motivation is very fast emulation and uncertainty quantification (UQ) for  simulators with functional outputs and large ensembles, where near real-time prediction is needed in experimental settings.

Gaussian process (GP) regression has become the gold standard for emulation \citep{Sacks89}. They are flexible predictors that quantify uncertainty in a principled way, but their cubic computational cost makes them impractical for large datasets. This challenge is compounded when outputs are functional, with dimension $d_y$. Treating each functional index as an input yields $Md_y$ scalar responses for an ensemble of size $M$, which is often infeasible. Efficient methods have been proposed which fit emulators to each of the $d_y$ indices \citep[e.g.,][]{robustgasp}, but can still become computationally intensive when $d_y$ is large \citep{hutchings_slosh}.

Orthogonal basis decompositions provide an effective alternative as complex and non-linear functional response can often be well represented by $p \ll d_y$ basis functions, reducing the number of response values to $Mp$. The seminal framework of \citep{KeOha01} was extended for functional response in this way by \citep{higdon}, fitting independent GPs to empirical orthogonal functions (EOFs) derived from a singular value decomposition (SVD). This approach, which we refer to throughout as SVDGP, has been widely applied \citep[e.g.,][]{higdon_flyer_plate}, but struggles to scale to large ensembles. Other dimension-reduction approaches such as active subspaces \citep{active_subspace,Ma_functional_em} have also been explored for GP models, but are still by cubic computational costs.

Several scalable GP frameworks have been developed to address this bottleneck, including local approximations \citep{gramacy_apley,laGP} and Vecchia-based methods \citep{Vecchia,sVecchia,katzfuss2021}. When combined with EOF decompositions, these methods can provide accurate emulation, but in practice neighborhood selection and parameter estimation can remain computationally demanding at the scale of tens of thousands of runs, which motivates the approach taken here.

In this work we propose a framework, termed FlaGP, that combines four ideas: (i) EOF decomposition for functional outputs, (ii) local GP emulation for scalability, (iii) input scaling based on global GP parameter estimates to improve neighbor selection, and (iv) modular Bayesian calibration. This combination yields an accurate and scalable emulation and calibration pipeline for large functional-output simulators, with the added benefit of enabling very fast active learning strategies.

These ideas extend prior work on emulation and modular calibration for scalar outputs \citep{GramacyBingham2015}. Our contributions are the extension to functional outputs, the integration of input scaling into local prediction, and the development of both Bayesian and maximum a posteriori (MAP) approaches to calibration. The remainder of this work is organized as follows. Section \ref{sec:Application} introduces the motivating application and dataset. Section \ref{sec:Emulation} gives more discussion on SVDGP and develops the proposed emulation framework. Section \ref{sec:working_ex} presents a working example comparing our method to SVDGP and an existing scalable GP approach. Section \ref{sec:seq_d} develops the active learning framework, while Section \ref{sec:calib} and \ref{sec:fast_pt_est} describe Bayesian and MAP calibration. Section \ref{sec:Al_example} analyzes the motivating dataset, and Section \ref{sec:discussion} concludes with a discussion of strengths and limitations.

\section{Application}\label{sec:Application}

The proposed methodology is demonstrated on the multiphysics continuum hydrodynamics code FLAG \citep{flag2}, a deterministic computer model for high-strain-rate and deformation responses for materials science. A large ensemble of $M=20000$ runs from FLAG is available for analysis, simulating an Aluminum sample impacted by an Aluminum flyer-plate over a range of experimental conditions. The ensembles span an 11-dimensional input space. Inputs include those for the Johnson-Cook strength model \citep{jcook} ($A$, $B$, $C$, $n_\text{JC}$, $m_\text{JC}$), the flyer impact velocity ($v_1$, $v_2$, $v_3$), and the material shear modulus ($G_1$, $\Delta_2$, $\Delta_3$). Simulator outputs are available for three ``shots" associated with varied flyer-plate and sample thickness; referred to as shots \textit{104}, \textit{105}, and \textit{106}. Each velocity and shear modulus parameter is associated with a single shot. 
\begin{figure}
    \centering
    \includegraphics[scale=.5]{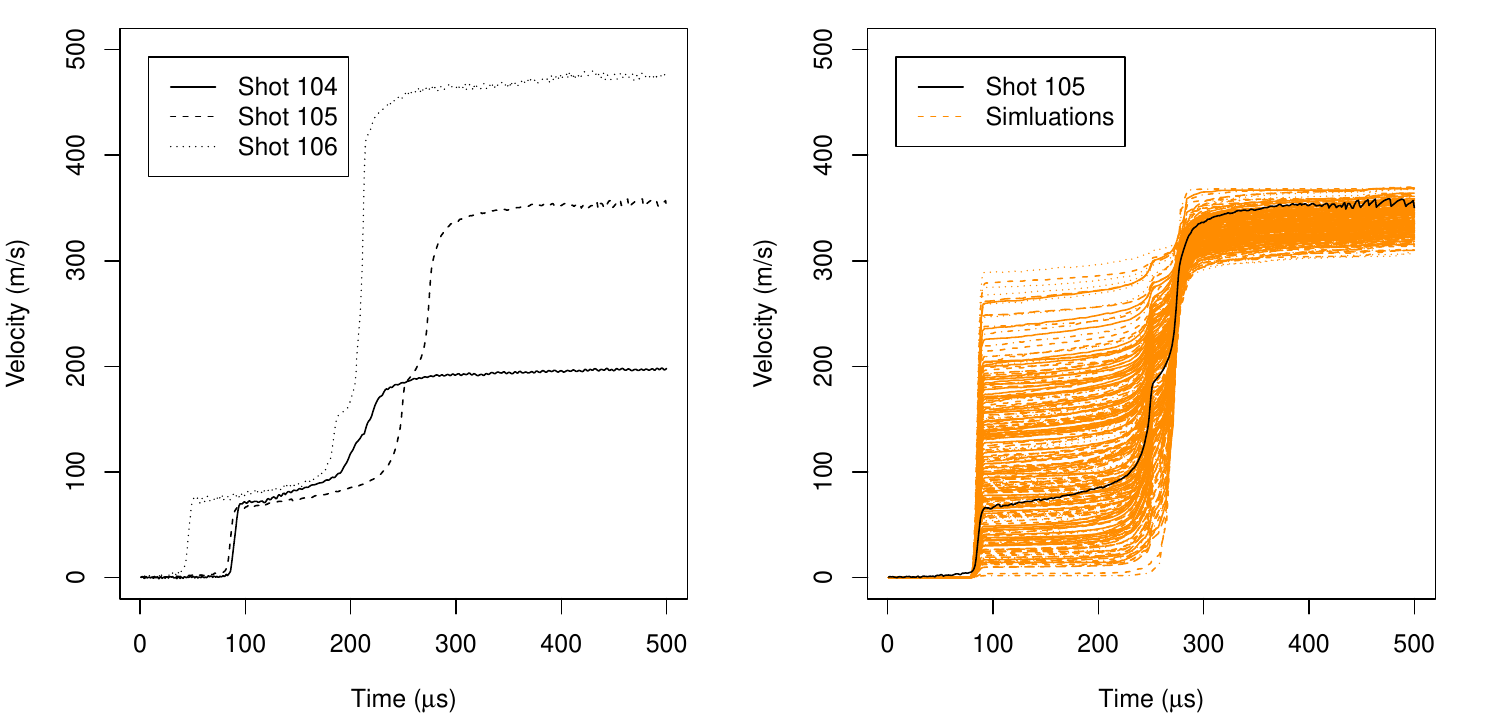}
    \caption{Left: Velocity profiles for three observed shots. Right: Shot \textit{105} and a subset of the FLAG simulations.}
    \label{fig:Alfeatures}
\end{figure}

The output of a simulation is a 1D profile of the impacted sample's free surface velocity over time. In addition to the computer model output, data is available from a single field experiment for each of the shots \citep{al5083experiments}. This work explores emulating FLAG as well as using the experiment to calibrate these 11 parameters. The aim is to deploy the emulator in an experimental facility where calibration can be done very quickly, enabling rapid updating with new experimental data. Fig.~\ref{fig:Alfeatures} shows the three experimental shots in the left panel, and on the right, shot \textit{105}, along with a subset of the simulations. 

A smaller ensemble of 1000 runs from FLAG, simulating the same experiments, was analyzed using the SVDGP model at great computational expense \citep{higdon_flyer_plate}. In their work, only 4 features along each curve were extracted for emulation and calibration. Section \ref{sec:Al_example} presents a computationally tractable analysis for this significantly larger ensemble that uses the entire velocity curve sampled at $d_y=200$ time points for each shot. The analysis of this larger dataset would be computationally infeasible with the SVDGP model. 

%Using a Python implementation of the SVDGP model in the \texttt{SEPIA} package \citep{SEPIA}, we estimate that their calibration took up to 10 days of computing time. The supplementary material contains a comparison to their analysis using FlaGP and we find that competitive emulation and calibrated predictions can be achieved in a fraction of the time. 

\section{Emulator Formulation}\label{sec:Emulation}

The common approach to computer model emulation for functional outputs is outlined here, with notation introduced and challenges for large datasets discussed.

Let $\bm{x} \in \mathbb{R}^{d_x}$ denote an input vector, and let $\bm{z}(\bm{x}) \in \mathbb{R}^{d_y}$ be the standardized computer model output at input $\bm{x}$. The $M$ input locations are the rows of the design matrix $\bm{X} \in \mathbb{R}^{M \times d_x}$ (scaled to the unit hypercube), and the corresponding model outputs in the matrix $\bm{Z}(\bm{X}) \in \mathbb{R}^{M \times d_y}$, whose $i$th row is $\bm{z}(\bm{x}_i)$. Outputs are standardized so that each output variable (each column of $\bm{Z}$) has 0 mean, and the total variance of all outputs is 1.

Following \cite{higdon}, we represent $\bm{Z}(\bm{X})$ via SVD:
\begin{equation}
    \bm{Z}(\bm{X}) = \bm{U} \bm{D} \bm{V}^T,
\end{equation}
where $\bm{U} \in \mathbb{R}^{M \times p}$ and $\bm{V} \in \mathbb{R}^{d_y \times p}$ are orthogonal matrices, $\bm{D} \in \mathbb{R}^{p \times p}$ is the matrix with singular values along the diagonal, and $p = \min(M, d_y)$.

We define the basis matrix $\bm{B} = \bm{U} \bm{D} / \sqrt{M}$, whose columns $\bm{b}_j$ provide orthogonal directions capturing dominant modes of variation in $\bm{Z}(\bm{X})$. The columns of $\bm{B}$ are ordered by the magnitude of the corresponding singular values in $\bm{D}$, so that the first column $\bm{b}_1$ captures the greatest variation in $\bm{Z}(\bm{X})$, the second column the next greatest, and so on. The associated weight matrix is $\bm{W}(\bm{X}) = \sqrt{M}\, \bm{V}^T \in \mathbb{R}^{p \times d_y}$.

We approximate the output as a truncated weighted sum of the first $p_\eta$ basis vectors ($p_\eta \ll p$), giving
\begin{equation}\label{eqn:em_form}
   \bm{\eta}(\bm{x}) = \sum_{j=1}^{p_\eta} \bm{b}_j\, w_j(\bm{x}),
\end{equation}
where $\bm{b}_j$ is the $j$th column of $\bm{B}$, and $w_j(\bm{x})$ is the weight for basis $j$ at input $\bm{x}$. At the training points, these weights are the entries of $\bm{W}$; for new input $\bm{x}^*$, they are predicted by Gaussian process (GP) regression defined by a mean function $\mu_j(\bm x)=\mathbb E(w_j(\bm x))$ and covariance $C(\bm x,\bm x')=\mathbb E\{[w_j(\bm x)-\mu_j(\bm x)][w_j(\bm x')-\mu_j(\bm x')]^\text{T}\}$. Generally, one works with the correlation matrix $K(\bm x,\bm x')=\sigma_{\text{gp}}^{-2}C(\bm x,\bm x')$, where $\sigma^2_\text{gp}$ is the variance. In this work, we use the isotropic Gaussian correlation function $K_{\bm l,g}(\bm x,\bm x')=\exp\{-||\bm x-\bm x'||^2/\bm l\} + g$ where $\bm l$ is the $d_x$-vector of \textit{lengthscale} parameters and $g$ is the \textit{nugget}.

By selecting only the first $p_\eta$ columns of $\bm{B}$, we retain the principal components that collectively account for the most variation in the outputs, yielding a low-dimensional representation. For the remainder of this work, we will simplify notation by using $\bm B$ and $\bm W$to refer to the first $p_{\eta}$ columns of $\bm B$ and rows of $\bm W$.

By modeling each weight function $w_j(\bm{x})$ as an independent GP, the high-dimensional emulation problem reduces to $p_\eta$ scalar GP regressions. In practice, $p_\eta$ is chosen so that the truncated representation $\bm{\eta}(\bm{X})$ approximates $\bm{Z}(\bm{X})$ with sufficiently small reconstruction error. This SVD-GP approach scales as $\mathcal{O}(M^3)$ operations, which limits its feasibility for very large datasets.

\subsection{Scalable Methodology}\label{sec:scalable}

In this section, we introduce the FlaGP emulator, which has the same functional form as \Cref{eqn:em_form}, but each $w_j(\bm x)$ is modeled using a modified laGP predictor. At a target input $\bm x$, laGP builds a neighborhood of $m \ll M$ training points--traditionally via a greedy search algorithm \citep{gramacy_apley}. While this reduces model fitting and prediction to $\mathcal O(m^3)$ operations, the greedy search step can itself become a computational bottleneck in settings like Bayesian calibration, where thousands of predictions must be made sequentially \citep{leapGP}. In Section SM3 of the supplementary material, we demonstrate this through a simulation study with dataset size similar to the application data. Furthermore, even though $m$ tends to be relatively small, the cost of estimating the parameters for each laGP prediction is not negligible in such settings.

To overcome these issues, the FlaGP emulator pre-estimates global GP lengthscale parameters and rescales each input by the inverse square root of its lengthscale. Scaling the input space allows for the circumvention of both lengthscale estimation for each laGP predictor, as well as the need for a greedy search to select training points. This will be discussed in detail in \Cref{sec:input_scaling}.

\subsubsection{Lengthscale Estimation and Input Scaling}\label{sec:input_scaling}

The proposed methodology uses input scaling \citep{hsu, satellite} to significantly reduce the computational burden of sequential prediction with laGP. An input $\bm{x}$ is scaled by estimates of the global lengthscale parameters $\hat{\bm{l}} = (\hat{l}_1,\ldots,\hat{l}_{d_x})$. That is, $\bm{x}^\text{sc} = \bm{x}/\sqrt{\hat{\bm{l}}}$, where vector division is element-wise.

Scaling inputs provides a twofold benefit in speed
\begin{itemize}
\item Nearest neighbors in the scaled input space are identical to those under a lengthscale-weighted Mahalanobis distance in the original space, which can dramatically improve the quality of NN designs for prediction. NN designs are much faster to compute than the near-optimal predictive neighborhoods implemented in laGP.
%point pairs with the highest \textcolor{blue}{prior?} correlation \textcolor{blue}{under the prior GP} correspond to nearest neighbors, which can dramatically improve the quality of NN designs for prediction when the GP correlation function is anisotropic. NN designs are much faster to compute than the optimal predictive neighborhoods implemented in laGP.
\item When predicting at $\bm x^{sc}$, lengthscale parameters do not need to be estimated because the global estimates are already included in the scaling. Lengthscales for models on scaled inputs $\bm x^{sc}$ are fixed at $\bm 1_{d_x}$.
\end{itemize} 

Input scaling in this way yields significant computational gains, but at the cost of sacrificing laGP's non-stationary flexibility. Our approach differs from that of \citep{satellite}, in that lengthscales used for prediction are fixed at $\bm 1_{d_x}$, and nearest neighbors in the scaled space are used, rather than initialized at $\bm 1_{d_x}$ to speed up parameter estimation.

Of course, the cost of pre-estimating lengthscale parameters cannot be ignored. In our implementation of the FlaGP framework, global lengthscale estimates are obtained using an empirical Bayes approach implemented in the \texttt{laGP} package. That approach uses Gamma prior distributions for the lengthscale parameters and MAP estimation with a subset of $m_\text{est}<M$ ensemble members. As in \cite{satellite}, bootstrapped block Latin hypercube sampling (BLHS) is used to build subsets for estimation, and the bootstrap median estimate is used. % which ensures consistency \citep{blhs}. 

%For high-dimensional input spaces like we have in the motivating examples ($d_x=11$), the computational effort of BLHS can be impractically high. For these applications, a simple stratified random sampling approach can be used, and repeated $r_\text{est}$ times to account for sampling variability. The median estimate over stratified samples can be used.

We have found that setting $m_\text{est}$ and the number of estimation replicates, $r_\text{est}$, as large as possible, given computational constraints, leads to improved estimation. While this increases the cost of precomputing, it does not affect the prediction and calibration costs of a deployed emulator. For FlaGP, $p_{\eta}$ sets of lengthscales, $\{\hat{\bm{l}}_1,\ldots,\hat{\bm{l}}_{p_{\eta}}\}$, must be estimated, one set for each of the independent components of the model. These are used to generate scaled inputs $\bm{X}^\text{sc}_1,\ldots,\bm{X}^\text{sc}_{p_{\eta}}$. Lengthscale estimation is only done once, and can be done in parallel, so the increased cost for $p_{\eta}$ estimations is often negligible.

%\textcolor{red}{The computational benefits of laGP with input scaling are clear, but it is important to note that it is a local predictor, not a global model for the data. Scalable GP frameworks based on the Vecchia approximation \citep{Vecchia} do provide a joint distribution for the data, but at a computational cost. If a global model for the data is not needed, the laGP platform with input scaling may provide a computational benefit, especially for tasks like Sequential Design which is discussed in \Cref{sec:seq_d}.} 

\subsubsection{Prediction}
Given lengthscale estimates, prediction at a new location $\bm{x}^*$ is made using transformed inputs $\bm{x}^\text{sc*}_j;\;j=1,\ldots,p_{\eta}$. Training data consists of the $m$ nearest neighbor inputs within $\bm{X}^\text{sc}_j$ with their associated weights within $\bm{w}_j$, which we define as $\bm{X}^\text{sc*}_j$ and $\bm{w}_j^*$. Integrating out the GP variance under a reference prior yields a Student-t predictive distribution \citep{gramacy_apley} with
\begin{equation}\label{eqn:w_lagp}
\begin{split}
    \text{mean} \;& \mu_j(\bm{x}^\text{sc*}_j)=\bm{c}(\bm{x}^\text{sc*}_j)'\bm{C}(\bm{X}^\text{sc*}_j)^{-1}\bm{w}^*_j,\\
    \text{scale} \;& s^2_j(\bm{x}^\text{sc*}_j)=\frac{\Psi_j(\bm X_j^\text{sc*})}{m}\big[1-\bm{c}(\bm{x}^\text{sc*}_j)'\bm{C}(\bm{X}^\text{sc*}_j)^{-1}\bm{c}(\bm{x}^\text{sc*}_j)\big],
%    w_j(\bm{x}^\text{sc*}|\mathcal{D}^*_j) \sim N(\mu_j(\bm{x}^\text{sc*}),\sigma^2_j(\bm{x}^\text{sc*}))\;j=1,\ldots,p_{\eta},
\end{split}
\end{equation}
and $m$ degrees of freedom. The $m$-vector $\bm{c}(\bm{x}^\text{sc*}_j)$ contains correlations between simulator outputs at $\bm{X}^\text{sc*}_j$ and the predicted output at $\bm{x}^\text{sc*}_j$. The $m \times m$ matrix $\bm{C}(\bm{X}^\text{sc*}_j)$ is a correlation matrix with $[\bm{C}]_{ll'} = C(\bm{X}^\text{sc*}_l,\bm{X}^\text{sc*}_{l'})$ and $\Psi_j(\bm{X}^\text{sc*}_j)=\bm{w}_j^{*T}\bm{C}(\bm{X}^\text{sc*}_j)^{-1}\bm{w}^*_j$. 

Letting $\bm{\mu}_{\bm{w}}=[\mu_{1},\ldots,\mu_{p_{\eta}}]^T$ and $\bm{s}^2_{\bm{w}}=[s^2_{1},\ldots,s^2_{p_{\eta}}]^T$, the predictive mean in the functional response space is $\bm{\mu}_{\bm{\eta}} = \bm{B}\bm{\mu}_{\bm{w}}$, and the predictive variance is $\bm{\Sigma}_{\bm{\eta}}=\bm{B}\bm{\Sigma_{\bm{w}}}\bm{B}^T$, where $\bm{\Sigma}_{\bm{w}} = \frac{m}{m-2}\texttt{diag}(\bm{s}^2_{\bm{w}})$. The predictive distribution of $\bm{\eta}(\bm{x^*})$ follows a multivariate Student-t distribution with mean $\bm{\mu}_{\bm{\eta}}$ and variance $\bm{\Sigma}_{\bm{\eta}}$, but in practice we often approximate it as $\bm{\eta}(\bm{x^*})\sim \text{N}(\bm{\mu}_{\bm{\eta}},\bm{\Sigma}_{\bm{\eta}})$ for computational reasons, which will be discussed in \Cref{sec:calib}. %Predictive samples of $\bm y(\bm x^*)$ are obtained by multiplying samples from the distribution defined by \Cref{eqn:w_lagp} by the first $p_{\eta}$ columns of the basis matrix $\bm{B}$. 
The emulation procedure is summarized below in \Cref{alg:em_form}. A more detailed version of the algorithm is given in Section SM1 of the supplementary material.

\begin{algorithm}
    \caption{Emulator Formulation}
    \begin{algorithmic}[1]
        \item \textit{Precomputing}: With inputs $\bm{X} \subseteq [0,1]^{d_x}$, %scaled to the unit hypercube,
        and standardized outputs $\bm{Z}$, %having mean zero and unit variance, 
        compute $\bm{B}=\bm{UD}/\sqrt{M}$ and $\bm{W}(\bm{X})=\sqrt{M}\bm{V}^T$ using the SVD $\bm{Z}=\bm{U}\bm{D}\bm{V}^T$.
        \item \textit{Inference}: Estimate correlation parameters $\hat{\bm{l}}_j;\;j=1,\ldots,p_{\eta}$ using the empirical Bayes procedure implemented in \texttt{laGP} with BLHS or stratified sampling. 
        \item \textit{Input Scaling}: Compute scaled inputs $\bm{X}^\text{sc}_j=\bm{X}/\sqrt{\hat{\bm{l}}_j};\;j=1,\ldots,p_{\eta}$.
        \item \textit{Prediction at $\bm{x}^*$}: For $j=1,\ldots,p_{\eta}$, find the $m$ nearest inputs $\bm{X}^\text{sc*}_j\subset \bm{X}_j^{sc}$, and their associated outputs $\bm w_j^*$, to the scaled input $\bm{x}^\text{sc*}_j=\bm{x}^*/\sqrt{\hat{\bm{l}}_j}$. The predictive distribution for each $\bm{x}^*_j$ is defined by \Cref{eqn:w_lagp}, and can be used to compute the native predictive mean and variance $\bm{\mu}_{\bm{\eta}}$ and $\bm{\Sigma}_{\bm{\eta}}$.
    \end{algorithmic}
    \label{alg:em_form}
\end{algorithm}

To make the application of the above steps clear, the proposed emulator is illustrated with a working example in Section \ref{sec:working_ex}, and its performance is compared to SVDGP. %This emulator formulation addresses only some of the computational challenges associated with fast functional computer model calibration. 
In Section \ref{sec:calib}, we show how to apply this emulator in a fast likelihood-based calibration framework.

\subsection{Working Example}\label{sec:working_ex}

In this section, a working example is developed which serves to demonstrate the viability of the FlaGP emulator and allows comparison to other state-of-the-art GP-based emulation techniques. For the working example, functional data are generated from a linear combination of cosine basis functions
\begin{equation}\label{eq:working_example}
    \bm y(\bm x,\bm\tau) = \sum_{p=0}^5 A(\bm x)cos(p \pi \bm\tau),
\end{equation}
where amplitude $A(\bm x)$ is a draw from a Gaussian process with $d_x=3$ inputs using a separable Gaussian correlation function. Lengthscales for the three inputs are drawn randomly from a uniform distribution on $[.05,.55]$. The functions are sampled at $d_y=50$ points $\bm \tau$ equally spaced on $[0,1]$.

A range of ensemble sizes from $M=100$ up to $M=10000$ is considered. This range of sample sizes serves to explore two important questions: (i) For ensembles where SVDGP can be implemented, how much better does it perform?; and (ii) for large ensembles where SVDGP is infeasible, does FlaGP perform similarly to other state-of-the-art approximate GP emulators? Choosing $M=10000$ is large enough to make comparisons to other scalable emulation methods, but not so large that we cannot make comparisons to SVDGP. In Section \ref{sec:Al_example} we will explore a more challenging application with $M=20000$. % and again compare laGP and sVecchia to the proposed emulator. %all emulation methods appear to be sufficiently converged at this point, so we do not consider larger ensembles.

\begin{figure}
    \centering
    \includegraphics[width=\linewidth]{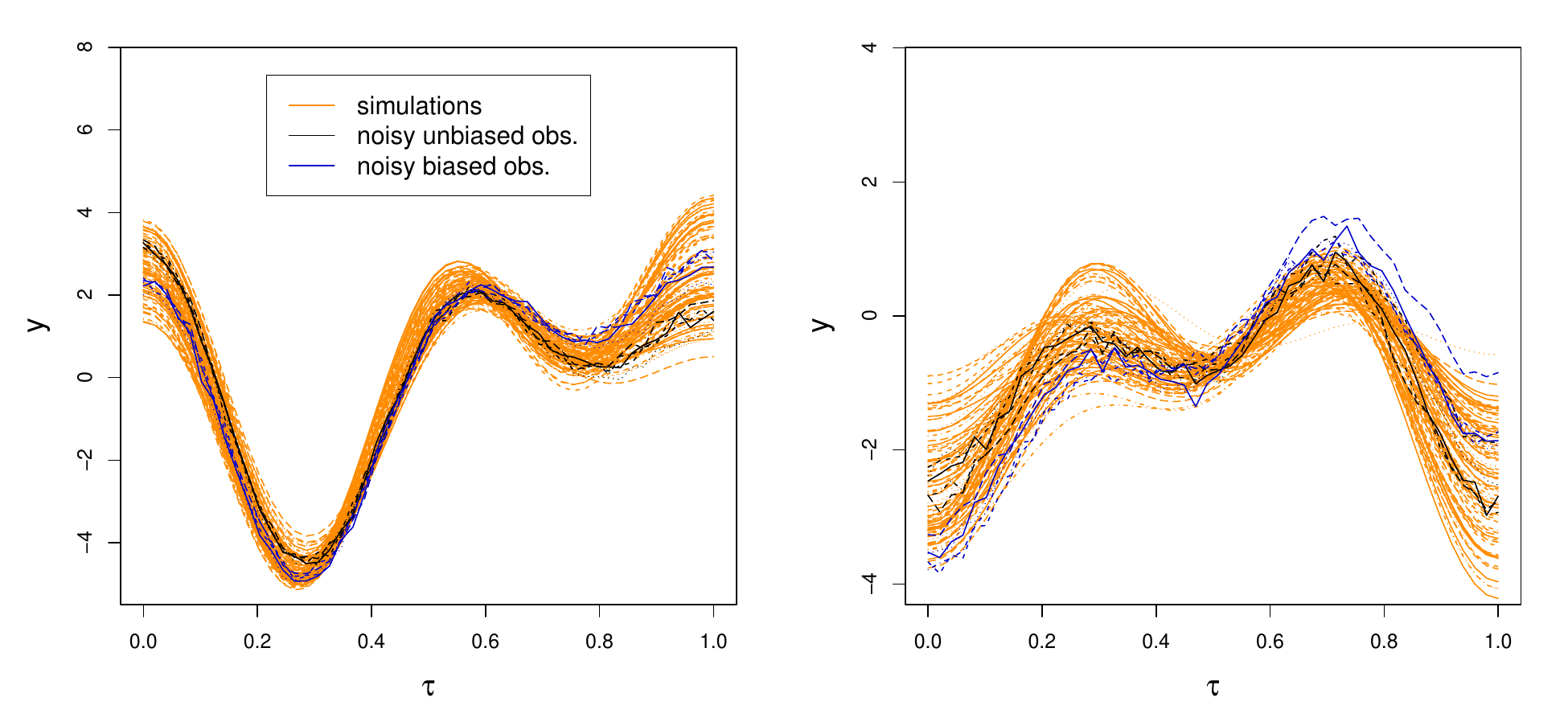}
    \caption{Example ensemble datasets generated from \Cref{eq:working_example}. $M=100$ simulations are shown in orange, which inform the modular emulator, 5 noisy experiments in black, and 5 biased noisy experiments in blue.}
    \label{fig:ensemble_vis}
\end{figure}

We generate 10 random datasets for each $M$, two of which are shown in \Cref{fig:ensemble_vis} for $M=100$. For now, we can ignore the black and blue curves shown in the figure, as they represent experimental data that will be used for calibration in \Cref{sec:unbiased_calibration,sec:biased_calibration}. %More details about the data generating mechanism can be found in Section SM\textcolor{red}{?} of the supplement.

The emulation methods considered for comparison are
\begin{enumerate}
    \item FlaGP as described in \Cref{sec:Emulation} (\url{https://github.com/granthutchings/FlaGP}).
    \item laGP with a separable Gaussian correlation function and the default ALC criterion for neighbor selection (implemented in the \texttt{R} package \texttt{laGP}). LaGP is trivially applied to functional response by parallelizing predictions over the $p_\eta$ components (\url{https://github.com/cran/laGP})
    \item The sVecchia model with a separable exponential correlation function, also applied in parallel over basis components (\url{https://github.com/katzfuss-group/scaledVecchia}).
    \item SVDGP using its default prior distributions, implemented using the \texttt{Python} package \texttt{SEPIA} (\url{https://github.com/lanl/SEPIA}).
\end{enumerate}

The methods selected for comparison are chosen for specific reasons. The proposed methodology is built upon the laGP framework and is specifically designed to tackle the computational issues associated with using laGP for Bayesian model calibration (i.e. speeding up sequential prediction), and thus it is natural to compare the proposed emulator with laGP. The sVecchia framework is selected because it also makes use of input scaling and has been shown to provide accurate emulation over a range of datasets \citep{sVecchia}.%, often outperforming Vecchia approximations without input scaling \citep{sVecchia}. %The proposed method combines aspects of both laGP and sVecchia, which, when combined, provide significant computational savings compared to either method.
Finally, we compare to SVDGP because it is the gold standard, using all the data at each step, and accounting for all parameter uncertainty in a Bayesian framework. We do our best to standardize the methods where possible to facilitate fair comparison, but some method-specific choices are required. 

\Cref{tbl:tuning_parameters} summarizes relevant tuning parameters for each approximate GP model over the ensemble sizes, as well as tuning parameters that are shared between all methods. Important inputs like the number of basis components, the nugget, and the neighborhood size are kept constant over the methods. Method-specific inputs like the starting neighborhood size for laGP and the size of the conditioning set for model fitting in sVecchia are kept fixed at their default settings.

\begin{table}
\begin{center}
\begin{tabular}{|c|c|c|c|c|c|c|c|}
\cline{3-8}
\multicolumn{2}{c|}{} & \multicolumn{6}{c|}{$M$} \\
\cline{3-8}
\multicolumn{2}{c|}{} & 100 & 250 & 500 & 1000 & 5000 & 10000 \\
\cline{3-8}
\hline
\multirow{2}{*}{\text{FlaGP}} & $d_\text{est}$ & 2 & 4 & 8 & 16 & 32 & 64  \\
\cline{2-8}
& $r_\text{est}$ & 5 & 5 & 5 & 5 & 5 & 5  \\
%\cline{2-8}
%&  &  &  & & & &  \\
\hline
\multirow{1}{*}{\text{laGP}} & start & 6 & 6 & 6 & 6 & 6 & 6  \\
%\cline{2-8}
%&  &  &  & & & &  \\
\hline
\multirow{1}{*}{\text{sVecchia}} & $m_\text{fit}$ & 30 & 30 & 30 & 30 & 30 & 30  \\
%\cline{2-8}
%&  &  &  & & & &  \\
\hline
\multirow{3}{*}{\text{All Models}} & $p_\eta$ & 2 & 3 & 4 & 5 & 5 & 5  \\
\cline{2-8}
& m & 10 & 25 & 50 & 100 & 100 & 100  \\
\cline{2-8}
& nugget ($g$) & $1e-5$ & $1e-5$ & $1e-5$ & $1e-5$ & $1e-5$ & $1e-5$ \\
\hline
\end{tabular}

\caption{Relevant tuning parameters for emulator comparison study.}
    \label{tbl:tuning_parameters} 
\end{center}
\end{table}

%For all models, the nugget parameter is fixed at $10^{-5}$ and the number of basis components is kept the same for all methods at each $M$. For the 6 ensemble sizes $2,3,4,5,5,$ and $5$ components are used, and is kept the same for each emulator. For FlaGP, laGP, and sVecchia, the number of data points used for prediction is kept the same at $m=\text{max}\{100,0.1M\}$. We found that increasing neighborhood size beyond 100 had no impact on predictive accuracy. 

%Lengthscale parameters for FlaGP are estimated using BLHS. To control fitting time, the number of divisions used in BLHS, denoted as $d_\text{est}$, is changed based on the ensemble size. For the 6 ensemble sizes $d_\text{est}= 2, 4, 8, 16, 32,$ and $64$ respectively. The number of bootstrap replicates is fixed at $r_\text{est}=5$ for all ensemble sizes. 

All model fitting and prediction for the four methods is done on the same computer, and all methods are parallelized over model components where possible, so computational comparison based on time is reasonable for the approximate GP methods. It is more difficult to make timing based comparisons with SVDGP because it is the only fully Bayesian method, and it is implemented in \texttt{Python} rather than \texttt{R}. %An MLE based implementation of a standard GP model was explored, but was still quickly infeasible for large $M$. 
For consistency with the Bayesian calibration study in \Cref{sec:calib}, the Bayesian implementation is used here, with 5000 MCMC samples collected after an initial tuning step. This is a modest number of samples, but we found that the models were well converged thanks to the initial MCMC tuning step.

%Results for the SVDGP model utilize the \texttt{Python} package \texttt{SEPIA} which implements an Bayesian GP using all the data. A Bayesian GP implementation using all the data is simply not computationally feasible for large ensemble sizes. For this comparison, we limit the ensemble sizes to $M=1000$. We do not find this to be a limitation for this simple working example because emulator accuracy quickly converges with increasing $M$ and these small ensemble sizes still provide insight into the different methods. As the ensemble size gets increasingly large, we would expect all the methods to perform similarly in accuracy.

%The laGP model is implemented through the \texttt{R} package \texttt{laGP} and \texttt{R} code to fit sVecchia is available at \url{https://github.com/katzfuss-group/scaledVecchia}. These 2 models are trivially applied to functional data by independently emulating each basis component in parallel. All model fitting and prediction is done on the same computer, so timing results are comparable.

FlaGP, laGP, and sVecchia use approximations that are useful for large datasets where standard GP approaches cannot be fit in a reasonable time. We therefore do not expect them to always perform as well in settings where SVDGP can be implemented. For the settings that we have explored, it seems that the approximate GP methods provide very competitive emulation accuracy and uncertainty quantification.

%The data for this working example are generated from a GP with 3 inputs $\bm x  \in [0,1]\times[0,1],\; t \in [0,1]$, which naturally limits the ensemble size to $M=1000$ given our ability to invert a large covariance matrix. Details for the data generating process are given in the supplementary material. 

%Fig.~\ref{fig:ub_C_data} shows $M=100$ evaluations over a single ensemble design as orange lines. Black dots and lines on this plot represent observed data for calibration and will be introduced in Section \ref{sec:unbiased_working_ex}.

%We explore the different emulators ability to accurately predict the functional output for a range of ensemble sizes. To account for the effect of a random ensemble on prediction, a Monte Carlo study is performed over 10 random ensemble designs for each ensemble size. For each design the proposed emulator is fit and predictions are made on a fixed test set of 1000 $(\bm x,t)$-pairs.% over the same domain as the ensemble designs. 

A fixed test set of 100 realizations from the data-generating mechanism is held out for out-of-sample predictive scoring. The quality of emulation is assessed via the root-mean-squared-error (RMSE) in the hold-out set. The quality of UQ is measured by the Interval Score (IS) \citep{scores}, an intuitive scoring rule favoring small intervals, but penalizing those that fail to include the true simulator output. The IS for confidence level $\alpha$ is defined as
$$\text{IS}(l,u|y,\alpha)=(u-l)+\frac{2}{\alpha}(l-y)\mathds{1}\{y<l\}+\frac{2}{\alpha}(y-u)\mathds{1}\{y>u\},$$
where $y$ is the true data value and $l,u$ are the lower and upper bounds of the prediction interval. IS is computed for level $\alpha=.05$ at each $\tau$ using the $2.5$ and $97.5$ quantiles of the predictive samples. %The score is averaged over all predictions.

\begin{figure}
    \centering
    \includegraphics[width=\linewidth]{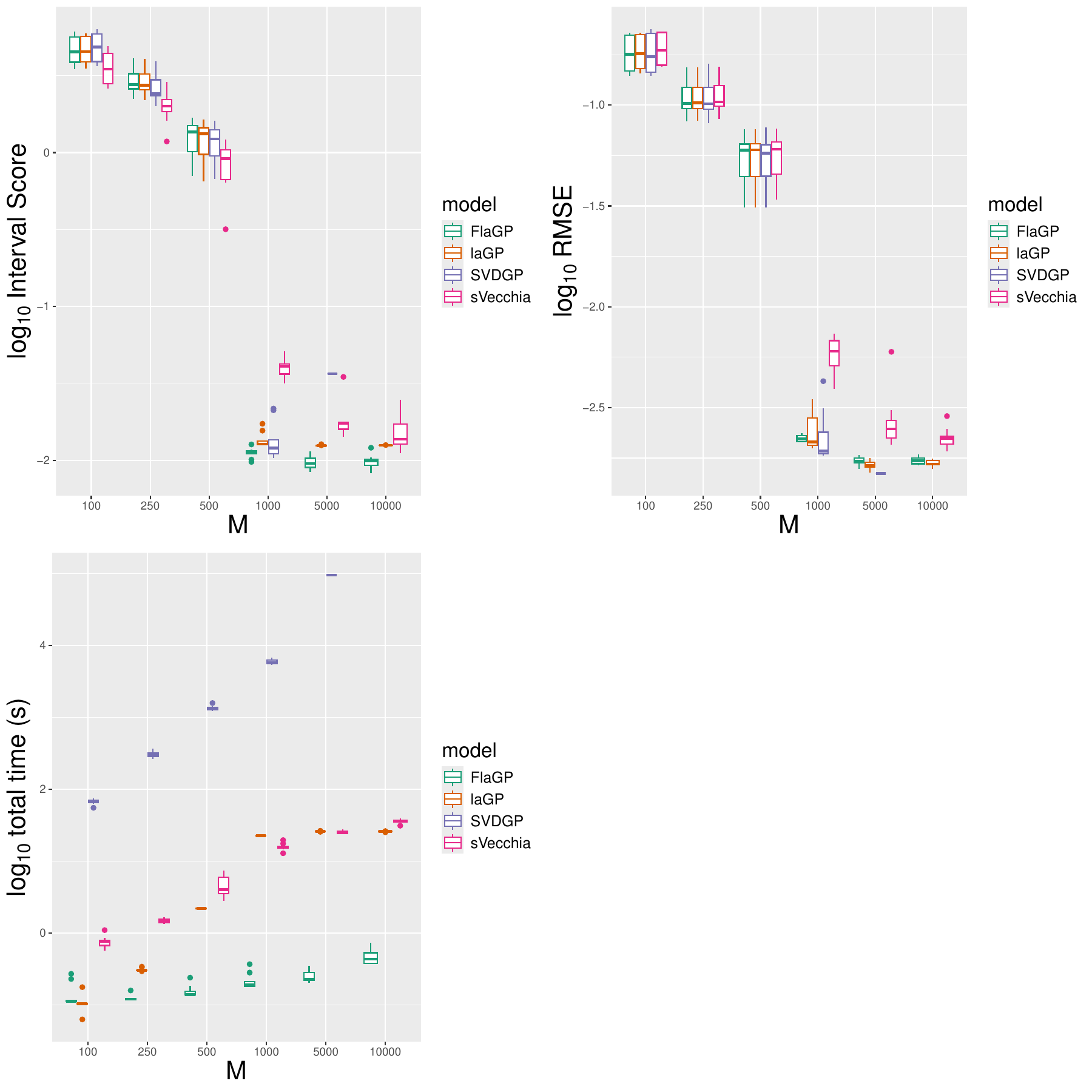}
    \caption{Predictive scoring and computation time for the working example. Results are given for 4 emulation methods over a range of ensemble sizes and 10 random ensemble designs. In the left panel, the logarithm of the RMSE on the test set is shown. In the center, the logarithm of the interval score for confidence level $\alpha=.05$. On the right, the total time (in log seconds) required for model fitting and prediction.}
    \label{fig:emulation_results}
\end{figure}

Box-pots of the log of the assessment metrics are presented over the random designs in \Cref{fig:emulation_results}. The top left panel shows the interval score, while the top right shows the RMSE. These scores assess both the accuracy and quality of uncertainty quantification in the predictions. In the Bottom left, we show the computation time for both model fitting and prediction. For SVDGP, we are computationally limited to $M=5000$, and there, only a single run of the 10 random ensembles is considered, requiring approximately 24 hours of compute time. %Comparison to SVDGP is limited to $M=1000$ due to the computational demands of the full GP model. %In Section \ref{sec:Al_example} a real data problem with $M=1000$ is explored where analysis with the full GP model is feasible but extremely computationally demanding.

 %We recognize that 1000 runs is not considered a large ensemble. %We explored, but do not present larger for this working example in part because predictive scoring is not notably improved past about $M=500$. 
%We do not find this to be a limitation for this simple example, as are still able to see the relative scaling with the ensemble sizes used, and we show that the proposed method is extremely fast compared to other state-of-the-art emulation methods.

%Predictive metrics are shown in \Cref{fig:emulation_results} where distributions are over the 10 random ensemble designs. 
Some important takeaways from the figure are
\begin{itemize}
    \item FlaGP gives prediction accuracy (top right) and UQ (top left) that is comparable to, or better than, the other methods, and is orders of magnitude faster for large $M$. Specifically, at $M=10000$, FlaGP required on average only $0.5$ seconds for fitting and prediction while laGP and sVecchia required 25 and 35 seconds, respectively.
    \item SVDGP tends to give very good prediction accuracy, and competitive UQ, but becomes computationally infeasible, requiring 24 hours for a single run at $M=5000$. If speed is important, any potential gains from SVDGP may not be worth the massive increase in computation time (bottom left).
    \item Generally, all four methods perform quite similarly in prediction accuracy and UQ. For small ensembles, sVecchia tends to provide the slightly improved UQ, likely due to its calibrated uncertainty method. However, for larger ensembles, -- the setting for which our method is designed -- the UQ from FlaGP may be preferable. 
    %\item FlaGP gives similar results to two methods designed for fast emulation of large ensemble datasets and proven on challenging application datasets \textcolor{red}{ref}.
\end{itemize}

For this simple working example, it seems that little to no concessions are made by selecting the fast approximate FlaGP methodology as compared to a full GP implementation like SVDGP, or another approximate GP method like sVecchia. To be clear, the difference in computation time between sVecchia and FlaGP is due to parameter estimation, not prediction. The sVecchia implementation uses a Fisher scoring algorithm by default, while our implementation of FlaGP uses BLHS, which can be much faster.

\section{Active Learning}\label{sec:seq_d}

Active learning is often used to maximally improve emulator performance for a given number of simulator runs by selecting an optimal input $\bm x_\text{c}$ to augment the current set of inputs $\bm X$, so that some performance metric is maximized (or minimized) over all possible inputs. Active learning has the potential to dramatically improve emulator performance, and the proposed method adapts readily to this approach. % for the same number of runs as compared to less targeted approaches like LHS \citep{mckay1992latin}. 

Improvement can be measured in many ways and may be problem-specific. For Bayesian optimization \citep{jones1998efficient}, the expected improvement is used to balance exploitation and exploration. That is, combining information from the model with a desire to explore the input space to select a new point intelligently. In the context of emulation, reducing the integrated mean squared prediction error (IMSPE) over the input domain $\mathcal X$ is a reasonable strategy. However, finding the point in $\mathcal X$ which gives the maximum reduction in IMSPE when added to the current ensemble $\bm X$ is a very computationally intensive continuous optimization problem which we solve using a discrete set of candidate points. 

The cost to solve this optimization problem scales exponentially with $d_x$ because of the number of points needed to approximate the IMSPE (an integral over $\mathcal X$). The set of candidate points $\bm X_c$ must be dense in $\mathcal X$ so that the selected point is near the true optimum. The IMPSE must be evaluated for $|\bm X_c|$ times, which, for a standard GP, makes this problem quickly untenable. Approximate GP methods facilitate significantly faster calculations of IMSPE due to fast prediction times.

%We show that even for a fast local predictor like laGP (and certainly for a standard GP), this optimization problem becomes quickly untenable due to the number of predictions required. 

In this section, we develop a fast active learning approach for FlaGP, which leverages its neighborhood structure to quickly determine which new design point from a set of candidate locations will most reduce the IMSPE of the emulator. We show that FlaGP allows for clever computation when calculating the IMSPE, leading to a significant reduction in computation compared to the default laGP.

IMSPE is not the only reasonable criterion for point selection in a active learning. We demonstrate how the much faster-to-compute and model-free maximin (MM) criterion can provide similar emulator improvement, as judged by out-of-sample predictive accuracy. The MM criterion simply selects the point from $\bm X_c$ that maximizes the minimum distance between the points in $\bm X_c$ and the current design $\bm X$. This space-filling metric ensures that no two design points are too close together. In addition to the traditional MM criterion, we develop a MM criterion for FlaGP which leverages the lengthscale estimates when computing distances between candidate and design points, encouraging the space filling property towards inputs which have the most impact on the response. The details of the (scaled) maximin metric will be discussed in \Cref{sec:scaled_mm}. 

%maximum variance (MV) criterion can provide very similar emulator improvements, as judged by out-of-sample predictive accuracy. The MV criterion simply selects the point from $\bm X_c$ with the maximum prediction variance, requiring only $|\bm X_c|$ predictions from the model. This is an intuitive metric that targets the regions of the input space where the model is most unsure, adding information there.

\subsection{IMSPE Criterion}

Here, we discuss the IMSPE criterion in more detail, which clarifies why IMSPE-based active learning is not feasible for a standard GP with large $M$ or $d_x$. IMSPE is defined as
$$I_{\bm X} = \int_{\mathcal X} \sigma^2(\bm x) d\bm x,$$
and can be estimated as the average prediction variance over a dense set of $n_I$ integration points in $\mathcal X$, that is,
$$\hat{I}_{\bm X} = \frac{1}{n_I}\sum_{i=1}^{n_I}\hat{\sigma}^2(\bm x_i).$$

%where $\hat{\sigma}^2(\bm x)$ is the computed prediction variance at $\bm x$.

%A common sequential design approach selects a candidate $\bm x_\text{c}$ which maximizes the reduction in integrated mean squared error (IMSPE) over $\mathcal X$ between the current ensemble $\bm X$ and the augmented ensemble $\bm X \cup \bm x_\text{c}$ \textcolor{red}{ref}. This can be a costly proposition for Gaussian process emulators because of the cost of computing the IMSPE over a large set of candidate locations. For a GP, we define the IMSPE under the design $\bm X$ as
%$$I_\bm X = \int_{\mathcal X} \sigma^2(\bm x) d\bm x,$$
%which can be estimated as the average prediction variance over a discrete set of integration points
%$$\hat{I}_\bm X = \frac{1}{n_I}\sum_{i=1}^{n_I}\hat{\sigma}^2(\bm x_i),$$ where $\hat{\sigma}^2(\bm x)$ is the computed prediction variance at $\bm x$. 

For a univariate GP with $M$ data points, computing $\hat{\sigma}^2(\bm{x})$ requires $\mathcal{O}(M^3)$ operations. For the functional setting, this is increased to $\mathcal{O}(p_{\eta}M^3)$ operations. The cost of computing $\hat{I}_{\bm{X}}$ is then $\mathcal{O}(n_I p_{\eta}M^3)$, and we cannot assume that $n_I \ll M$ because $n_I$ scales exponentially in $d_x$ as the set of integration points must be dense in $\mathcal{X}$ for a good approximation. Selecting a new design point requires computing $\hat{I}_{\bm{X} \cup \bm x_\text{c}}$ over a dense set of $n_c$ candidate points, and the total cost to select a single new point is $\mathcal{O}(n_c n_I p_{\eta}M^3)$ which becomes quickly infeasible as $M$ and/or $d_x$ increases. %In this section, we show that the scaled nearest neighbor approach of FlaGP enables IMSPE-based sequential design at a significantly reduced computational cost. %For the remainder of this section we will drop the hat notation and simply refer to the estimated IMSPE under design $\bm X$ as $I_\bm X$.

%The candidate input is selected to maximize $f(\bm x_\text{c}) = \text{IMSPE}({\bm X}) - \text{IMSPE}({\bm x_\text{c}})$ which gives the point that provides the greatest reduction in IMSPE over the input space when added to the ensemble. Since we cannot compute $f(\bm x_\text{c})$ for all possible candidates, we select the candidate that maximizes $f$ over a discrete set of $n_{cand}$ points in $\mathcal X$. Sequential design can be such a costly proposition because the number if integration points $n_{int}$ must be sufficiently large in order to approximate the IMSPE accurately for any given candidate location, and the number of candidate points $n_{cand}$ must be sufficiently large to ensure that the selected candidate is near to the point which provides the global maximum of $f$. Both $n_{int}$ and $n_{cand}$ should grow exponentially with the dimension of the input space. For a Gaussian process emulator, the computational cost of selecting a single point is $\mathcal O(n_{int}n_{cand}M^3)$.

%The proposed emulator can significantly reduce the cost selecting a new input. 

%Without any clever implementation,
The proposed emulator reduces the complexity of computing $\hat{I}_{\bm{X}}$ to $\mathcal{O}(n_cn_Ip_{\eta}m^3)$ simply by leveraging local prediction. However, we can further reduce this cost. If $\hat{\sigma}^2(\bm x)$ has been calculated for all $\bm x \in \bm X_\text{I}$, then to evaluate the updated $\hat{I}_{\bm{X} \cup \bm{x}_\text{c}}$ for a candidate design point, one only needs to recompute $\hat{\sigma}^2(\bm{x})$ for those $\bm{x}$ where the prediction neighborhood is effected by $\bm{x}_\text{c}$. For those integration points where the candidate point will not be included in its neighborhood, the prediction variance for the original model and the augmented model will be identical. The number of points for which $\hat{\sigma}^2(\bm{x})$ must be recalculated can be significantly fewer than $n_I$, further reducing computational cost.

For functional response models using a basis decomposition, the situation is slightly more complex. There are $p_\eta$ GP models, each with its own unique scaled input space. The set of integration points for which the predictive variance must be updated is the union of the set of points that are affected by $\bm{x}_\text{c}$ over all $p_\eta$ models. This set can still contain far fewer than $n_I$ points, but it begs an interesting question: are the points selected using only the first or first few basis vectors sufficient for emulator improvement? It may be that significant computational savings can be had by selecting the point that gives the most improvement to only the first GP (which accounts for the most variance in the data). The answer to this question will inevitably be problem-specific and likely depends on the variance proportion accounted for in the first few components. We do not explore this question systematically in this work, but we do present some results for the working example of \Cref{sec:working_ex}, which indicate that using only the first component may be sufficient for some applications. %Note that for our working example, the first weighted basis function captures 56\%-89\% of the variance in the data, a large proportion.

\subsection{Scaled Maximin Criteria}\label{sec:scaled_mm}

Computing the next optimal point under a standard maximin criterion is straightforward and model-free. Given a current design $\bm X$, and a set of candidate inputs $\bm X_c$, compute the distance matrix $D_c$, which contains, for each candidate $\bm x_c \in \bm X_c$, the distance from $\bm x_c$ to every point in $\bm X$. Maximin selects the $\bm x_c$ for which the minimum distance between $\bm x_c$ and all points in $\bm X$ is maximized. This is relatively fast to compute, compared to IMSPE, model-free, and gives a space-filling design in each dimension.

It is reasonable to assume that maximin could be improved by incorporating global lengthscale estimates to encourage the space-filling property towards important dimensions for the response. We define the scaled-MM criterion, which uses distance matrices $D^{sc}_j$, based on scaled inputs $\bm X^{sc}_j$ and scaled candidates $\bm X_{c,j}^{sc}$ for each basis component $j=1,\ldots,p_{\eta}$. Each distance matrix can be naturally weighted according to that basis components singular value $\lambda_j$, giving a weighted distance matrix \[D^{sc}=\sum_j \frac{\lambda_j}{\sum_k \lambda_k}D^{sc}_j.\]

The next optimal point in a sequential design procedure can be computed in maximin fashion according to the weighted distance $D^{sc}$. As with IMSPE, the scaled-MM criterion can be computed using fewer than $p_{\eta}$ basis components. However, there are less computational gains to be made by doing so unless $\bm X$ or $\bm X_c$ is large enough that computing distance matrices is prohibitive, and those calculations cannot be parallelized over basis components.

\subsection{Working Example}\label{sec:seq_d_working_ex}

This section investigates the effect of (i) the metric used for selecting candidate points (IMSPE, MM, or scaled-MM), and (ii) the number of model components used for candidate selection. These effects are explored for the example presented in \Cref{sec:working_ex}. Emulators are initialized with $M=9$ design points, one point near each of the 8 corners of the 3-dimensional design space, and one central point. New design points are added iteratively from a set of 500 possible candidates, and the IMSPE is evaluated on a set of 1000 integration points, selected using a maximin LHS.  The process continues until each emulator has $M=100$ total points. The small initial and final $M$ are used for illustrative purposes, and the main conclusions hold for larger datasets.

We compare our fast active learning procedure using FlaGP to laGP, further illustrating the computational savings from the proposed framework. For both FlaGP and laGP, the number of neighbors used for prediction is fixed at $m=\text{min}\{M,20\}$, and the number of components used for each model is $p_\eta=5$. The FlaGP emulators are refit (updating basis vectors and parameter estimates) every 10 new design points, and laGP uses these updated basis vectors.

\begin{figure}
    \centering
    \includegraphics[width=\linewidth]{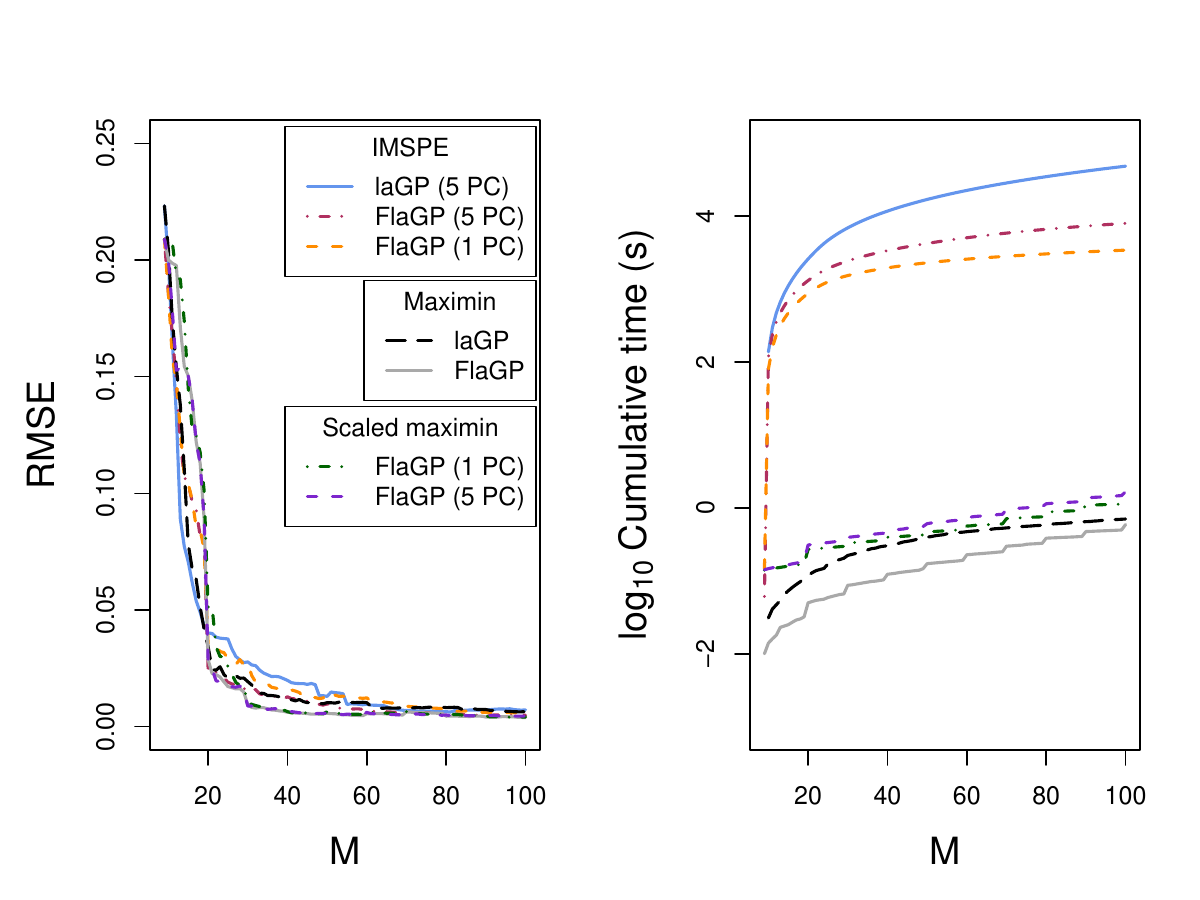}
    \caption{Test set RMSE (left) and cumulative computation time (right) plotted against ensemble size for the active learning working example. Points are selected via active learning using the (scaled) maximin and integrated mean squared prediction error criteria. For these data, there is no benefit to the expensive IMSPE-based design approach. Jumps in computation time for FlaGP are due to model refitting every time 10 new points are added to the design.}
    \label{fig:sequential_design}
\end{figure}

The effectiveness of the active learning procedure is evaluated using the RMSE (as in \cref{sec:working_ex}) on a hold-out set of 1000 evaluations from the model, none of which are included in the initial design, or the candidate set. The RMSE and cumulative compute time for both point selection criteria are plotted against $M$ in \Cref{fig:sequential_design}. As we discussed, using a reduced number of model components to evaluate the selection metric can provide significant savings for FlaGP. We consider selection based on only the first component, as well as all 5 components.

\Cref{fig:sequential_design} illustrates that significantly more computation is required for the IMSPE criterion, but using only the first model component for point selection can significantly reduce that cost. Using the IMSPE criterion with all 5 model components, FlaGP required 2.2 hours, but using (scaled) maximin, at most 1.5 seconds. If only 1 basis component is used with FlaGP, the cost for IMSPE is reduced to under 1 hour. In contrast, laGP is much more expensive; using the IMSPE criterion, laGP with all basis components required over 13 hours. Note that jumps in computation time for FlaGP are due to model refitting every time 10 new points are added.

For this example, the difference between methods' predictive accuracy is practically insignificant. The IMSPE criterion does not provide an improvement over MM for these data, but we suspect that this will not be the case for all computer models, and some will benefit greatly from the more targeted active learning approach. For example, we found that there may be some improvement for the Aluminum flyer-plate dataset, which presents a more difficult emulation task with moderate dimensional inputs and outputs. Active learning for the Aluminum flyer-plate data is explored in \Cref{sec:al_active_learning}. Importantly, the difference in computation time is very significant for this example, and for the Aluminum data.

These results warrant further exploration, especially the effect of the number of model basis components used for the selection criterion. For these data, using only 1 model component gave similarly effective designs, but for scaled-MM, this comes at a notably reduced computational cost. This could be for a number of reasons. It may be due to the large proportion of variability explained by the first component ($\approx 60\%$), or it could be simply due to the complexity of the emulation task. For this example, we suspect the latter because the basic maximin approach is very effective.

This example serves to illustrate two important points;
\begin{enumerate}
\item FlaGP facilitates faster IMSPE-based active learning than laGP by leveraging scaled nearest neighbor prediction. With FlaGP, the computational cost can be further reduced using a reduced number of model components in the selection criterion.
\item For large ensembles and high-dimensional input spaces, (scaled-)MM should be considered as a point selection criterion.
\end{enumerate}

A keen observer may have noticed that RMSE occasionally increases slightly with ensemble size in \Cref{fig:sequential_design}. When using nearest neighbors, replacing a point in the local prediction neighborhood--which requires removing the current furthest point in the neighborhood--does not always reduce prediction error. A simple example in Section SM 2 shows this can be common for GP models due to changes in the geometry of the neighborhood.

\section{Modular Calibration}\label{sec:calib}

An important use case for emulators is model calibration. In calibration problems, there are two types of computer model inputs: control inputs $\bm{x}$, which are measurable by the experimenter, and calibration inputs $\bm{t}$. Calibration inputs are set for the simulation trials, but the value for the field data, $\bm t=\bm{\theta}$, is unknown and must be estimated. The goal of computer model calibration is to use field observations and the computer model to (i) construct a predictive model for the system with uncertainty; and (ii) estimate $\bm{\theta}$. This section describes how the FlaGP emulator of Section \ref{sec:Emulation} can be used for fast model calibration.

Unlike \cite{KeOha01} and \cite{higdon}, the proposed method takes a modular approach to calibration \citep{modular}, where the emulator is fit using only the computer model data. Doing so can considerably speed up calibration since the FlaGP model is precomputed (see the previous section). This approach is especially relevant for the types of problems motivating this work, which tend to have both small $n$ and large $M$. In this case, %fitting the emulator jointly to both the computer model output and observed data adds computational complexity, but 
often little is to be gained by the addition of a mere handful of field observations field observations to inform the emulator.

In some cases, the computer model is known to systematically deviate from observations. The common approach is to augment the statistical model with a discrepancy term to account for the systematic bias. Calibration with and without discrepancy requires slightly different formulations. In Section \ref{sec:unbiased_calibration}, a calibration framework is developed for the no-discrepancy case. In Section \ref{sec:biased_calibration}, we discuss the addition of a discrepancy model.

Issues can arise in computer model calibration when the empirical bases, $\bm{B}$, do not represent the observations well. In \cite{salter} an algorithm is proposed to rotate $\bm{B}$ such that the reconstruction error for the observations is minimized. For FlaGP, this can be added as a precomputing step if needed. For the examples presented in this section, reconstruction error is not an issue, and for our large application datasets (e.g. \Cref{sec:Al_example}), we found these methods to be computationally prohibitive.

\subsection{Unbiased Calibration}\label{sec:unbiased_calibration}

Consider the setting where there are $n$ noisy field observations, each parameterized by a vector of control inputs $\bm{x}^F_i=(x_{i1},\dots,x_{id_x})^T$ with a corresponding functional response $\bm{y}(\bm{x}^F_i) \in \mathbb{R}^{d_y};\;i=1,\ldots,n$. The computer model $\bm{\eta}(\bm{x},\bm{t}) \in \mathbb{R}^{d_{\eta}}$ depends not only on $\bm{x}$, but also on the calibration inputs $\bm{t}=(t_1,\ldots,t_{d_t})$. With the assumption that the observed data is a noisy version of the simulator output, the model for the field data is 
\begin{equation}\label{eqn:ub_model}
    \bm{y}(\bm{x}_i^F) = \bm{\eta}(\bm{x}_i^F,\bm{\theta}) + \bm{\epsilon};\;\bm{\epsilon} \sim N(\bm{0},\sigma^2_y\bm{I}_{d_y}).
\end{equation}

If $\bm \eta$ is as defined in \Cref{sec:Emulation}, and the Gaussian approximation to the Student-t predictive distribution is used, the vector of observations follows the conditional Gaussian distribution
\begin{equation}\label{eqn:func_y_dist}
    \bm{y}(\bm{x}_i^F)\;|\bm\mu_{\bm \eta}(\bm x_i^F,\bm\theta),\bm \Sigma_{\bm \eta}(\bm x_i^F,\bm\theta),\bm{\theta},\sigma^2_y \sim N(\bm\mu_{\bm \eta}(\bm x_i^F,\bm\theta),\bm \Sigma_{\bm \eta}(\bm x_i^F,\bm\theta) + \sigma^2_y\bm{I}_{d_y});\;i=1,\ldots,n,
\end{equation}
where $\bm\mu_{\bm \eta}(\bm x_i^F,\bm\theta)=\bm \mu_{\bm w}(\bm x_i^F,\bm\theta)\bm B$ and $\bm \Sigma_{\bm \eta}(\bm x_i^F,\bm\theta) = \bm B^T\bm \Sigma_{\bm w}(\bm x_i^F,\bm\theta)\bm B$ are the projected posterior mean and covariance matrix of the emulator at inputs $(\bm x_i^F,\bm\theta)$. Each experiment is conditionally independent given the emulator, so the joint likelihood is the product of the individual likelihoods. The covariance matrix $\bm\Sigma_y=\bm \Sigma_{\bm \eta} + \sigma^2_y\bm{I}_{d_y}$ is $d_y \times d_y$, and the inversion of this matrix can be a bottleneck for likelihood evaluation for large $d_y$. In practice, this is not a limiting factor for our implementation as the likelihood can be evaluated in $\mathcal{O}((n(p_{\eta}+p_{\delta}))^3)$ operations, rather than $\mathcal O(d_y^3)$ operations by applying $\textit{Result 1.}$ from \cite{higdon}. This will be discussed in further detail in \Cref{sec:fast_pt_est}. For the problems we tend to encounter, observed data is limited, and $n(p_\eta+p_\delta)\ll 1000$, so likelihood evaluation is fast.

In contrast to the lengthscale estimation procedure described in \Cref{sec:input_scaling}, for the calibration parameters $\bm{\theta}$, we sample from a Bayesian posterior distribution. The likelihood function depends on both $\bm{\theta}$ and $\sigma^2_y$. By default, we assign independent uniform prior distributions on the interval $[0,1]$ to each $\theta_k$ for $k = 1, \ldots, d_t$ and an Inverse-Gamma prior distribution for $\sigma^2_y$ with hyperparameters $\alpha_p = 1$ and $\beta_p = 0.001$. These prior choices are made to closely match the default priors used in \cite{SEPIA}, the \texttt{Python} implementation of the SVDGP framework, facilitating fair comparison.

The unnormalized posterior
\begin{equation}\label{eqn:func_post}
    p(\bm{\theta},\sigma^2_y|\cdot) \propto \prod_{i=1}^n p(\bm{y}(\bm{x}_i^F)\;|\;\bm{\mu}_{\bm{\eta}}(\bm{x}_i^F,\bm{\theta}),\bm{\Sigma}_{\bm{\eta}}(\bm{x}_i^F,\bm{\theta}),\bm{\theta},\sigma^2_y) \times p(\bm{\theta}) \times p(\sigma^2_y)
\end{equation}
can be sampled using MCMC. 

\Cref{alg:ub_calib} illustrates how to quickly sample from the unnormalized posterior distribution in \Cref{eqn:func_post} for inference on $\bm{\theta}$ and $\sigma^2_y$. %to compute $\bm{\mu}_{\bm{\eta}}(\bm{x}_i^F,\bm{\theta})$ and $\bm{\Sigma}_{\bm{\eta}}(\bm{x}_i^F,\bm{\theta})$ to evaluate the %the emulator of Section \ref{sec:Emulation} is used to quickly evaluate the 
%RHS of \Cref{eqn:func_post} for a specific $\bm{\theta}$ and $\sigma^2_y$.
\begin{algorithm}
    \caption{Calibration without a discrepancy model}\label{alg:ub_calib}
    \begin{algorithmic}[1]
        \item \textit{Precompute FlaGP Emulator}: Steps $1,2,$ and $3$ of \Cref{alg:em_form}
        \item \textit{Sample parameters}: Given a current value of the parameters, $\bm{\psi}_{\text{curr}} = [logit(\bm{\theta}),log(\sigma^2_y)]$, sample a new value $\bm{\psi}_{\text{new}}$. Since $\bm{\theta}\in[0,1]^{d_x}$ and $\sigma^2_y>0$, we use a multivariate normal distribution to jointly propose $\bm{\psi}_{\text{new}} \sim \text{N}(\bm{\psi}_{\text{curr}},\bm{\Sigma}_{\text{prop}})$
        \item \textit{Predict Observations}: Use the scaled nearest neighbor strategy in step 4 of \Cref{alg:em_form}, compute $\bm{\mu}_{\bm{\eta}}(\bm{x}_i^F,\bm{\theta})$ and $\bm{\Sigma_{\bm{\eta}}}(\bm{x}_i^F,\bm{\theta})$ for $i=1,\ldots,n$.
        \item \textit{Log posterior density}: Evaluate \Cref{eqn:func_post} given the emulator mean and variance, and sampled $\bm{\theta}$ and $\sigma^2_y$. The likelihood is computed from the conditional normal distribution defined in \Cref{eqn:func_y_dist}.
        \item \textit{Repeat sampling}: Repeat steps $2-4$ in a Metropolis-Hastings MCMC (MH-MCMC) sampler for posterior inference on $\bm{\theta}$ and $\sigma^2_y$.
    \end{algorithmic}
\end{algorithm}
We found that the adaptive covariance scheme proposed by \cite{adaptiveJointMCMC} works very well in a variety of test problems. In our implementation, the proposal covariance is adapted only during an initial burn-in phase so that samples used for convergence diagnostics and prediction use a constant proposal covariance matrix. %chain is initialized with a diagonal covariance matrix and updated every 50 iterations. The proposal covariance is adapted only during an initial burn-in phase so that samples used for convergence diagnostics and prediction use a constant proposal covariance matrix.

\subsubsection{Working Example}\label{sec:unbiased_working_ex}

%Returning to the working example, $n=5$ noisy simulated field observations are generated from \Cref{eqn:ub_datagen}, and the parameter $t$ is used as a calibration input. Simulated field observations are generated using $\;x \in \{-4,-2,0,2,4\}$ and $t=1$, with random errors $\bm{\epsilon} \sim N(0,.05^2)$. %The error variance was chosen to be approximately $5\%$ of the variance in the simulations at $d=5$ meters over a range of $M$. 
%Fig.~\ref{fig:ub_C_data} shows the observed data along with a representative ensemble of size $M=100$. %Computer model data is decomposed into bases and weights just as in Section \ref{sec:working_ex}. 
%This example illustrates the viability of the proposed calibration methodology in both speed and accuracy. 

Returning to the working example, we demonstrate fast Bayesian model calibration with FlaGP under the assumption that the simulator is an unbiased representation of the expected value of the observations (i.e., no discrepancy model is required when calibrating the emulator).

For calibration problems, the FlaGP model can be applied with two different neighborhood sizes, one for prediction during MCMC and the other for calibrated predictions with the fitted model. The neighborhood size used for MCMC can be a limiting factor for computation, and may need to be kept smaller, while a larger neighborhood is likely affordable for calibrated predictions. Define $m_c$ to be the neighborhood size used for prediction during the calibration phase, and $m$ the neighborhood size used for calibrated predictions. For this example, we let $m$ be as defined in \Cref{tbl:tuning_parameters}, and keep $m_c=20$ fixed for all $M$, which facilitates very fast calibration.

Emulators are fit using the same ensemble designs as in Section \ref{sec:working_ex}, with the caveat that two of the three inputs are now calibration inputs, so we have $d_x=1$ and $d_t=2$. For each ensemble, $n=5$ noisy observations are generated from \Cref{eq:working_example} with random errors drawn from a $\text{N}(0,.1^2)$ and added to each element of the functional response. For each ensemble, all 5 observations are generated with the same $\bm \theta$, drawn uniformly from $[0,1]\times[0,1]$ and $x$ varied over $[0,1]$. Examples of these noisy observations are shown as black lines in \Cref{fig:ensemble_vis}.

Comparison to the SVDGP model is made to illustrate the competitive parameter calibration (and calibrated predictions) of FlaGP with significantly reduced computational effort, and the ability to tackle calibration problems for very large $M$ where SVDGP is infeasible. For both FlaGP and SVDGP, 5000 posterior samples of the calibration parameter are generated. Comparison is not made with other models, such as laGP and sVecchia, because they do not implement a functional response calibration framework. %Implementing these calibration methods would require significant investigator effort, which is beyond the scope of this work. 
Furthermore, we would not expect calibration under these models to be a notable improvement over FlaGP, given the results of \Cref{sec:working_ex}.

\begin{figure}
    \centering
    \includegraphics[width=0.75\linewidth]{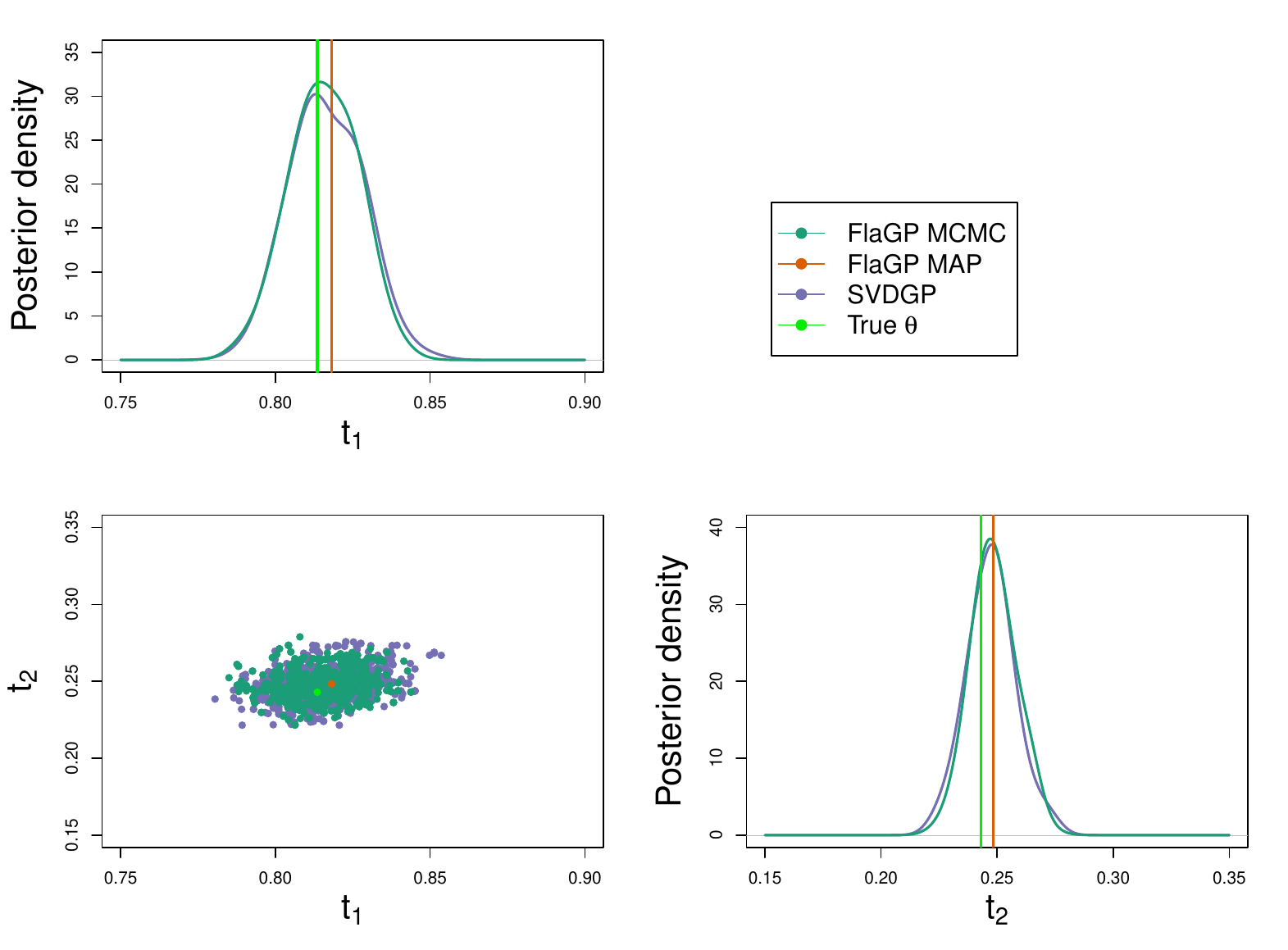}
    \caption{Marginal posterior density estimates and bivariate samples for the unbiased calibration working example from 1000 post-burn-in MCMC samples.}
    \label{fig:calib_bv_density}
\end{figure}

Since there is no discrepancy model, and the observed data are generated from the same process as the simulations, we expect our posterior distributions to have a mode near the true $\bm \theta$. \Cref{fig:calib_bv_density} shows posterior density estimates from 1000 post-burn-in samples and how they compare to the true $\bm{\theta}$ for a single ensemble dataset. Both FlaGP and SVDGP produce similar posterior distributions, with the true $\bm{\theta}$ being very close to the posterior mode.

The quality of calibrated predictions is assessed on a hold-out set of 100 noisy observations generated in the same way as the field observations, that is, the same $\bm \theta$, and $x$ varied over $[0,1]$. For both FlaGP and SVDGP, 100 posterior samples are used for prediction, and those predictions are assessed using RMSE, IS, and empirical coverage probability.

\begin{figure}
    \centering
    \includegraphics[width=\linewidth]{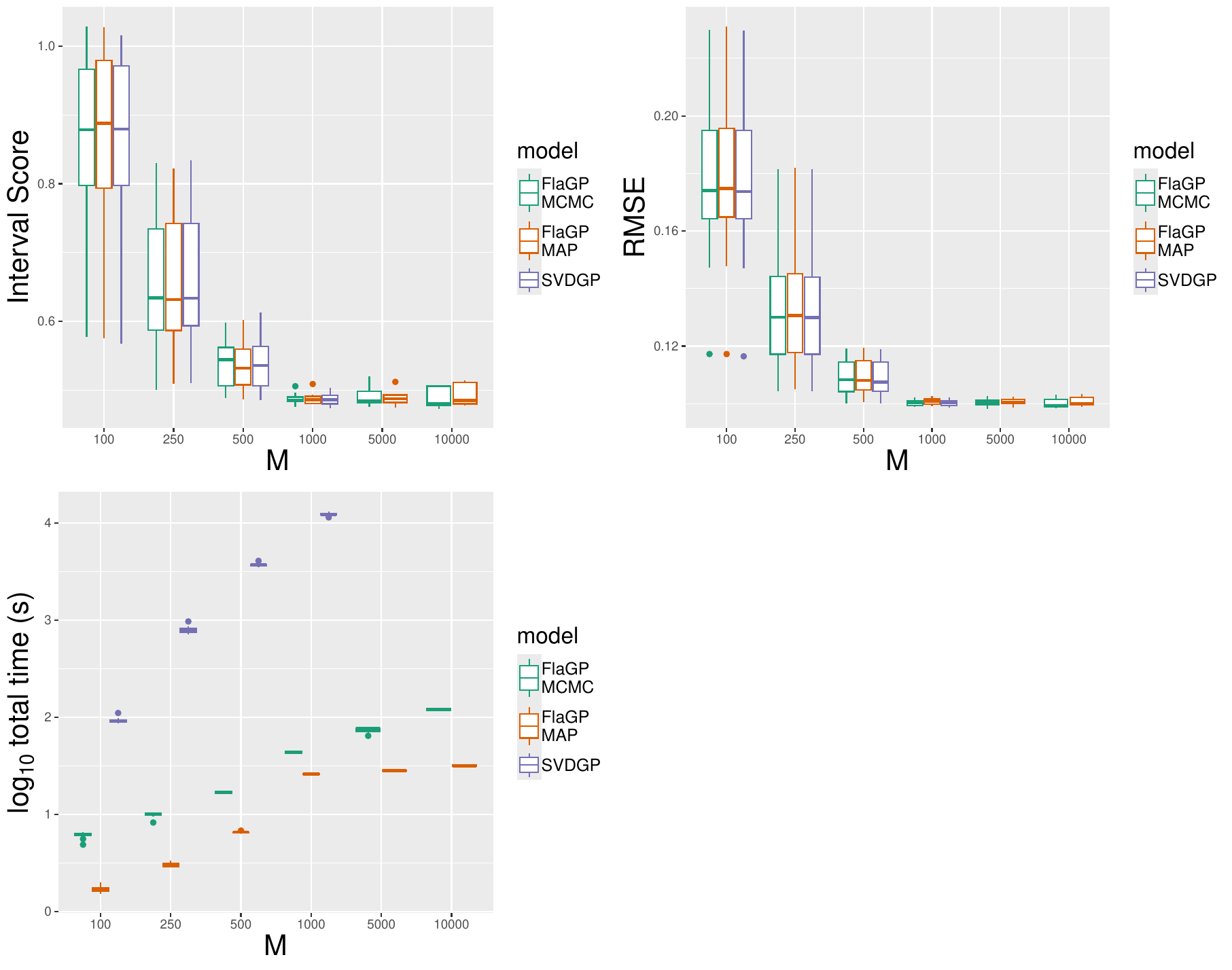}
    \caption{Predictive scoring on the test set and computation time for the unbiased calibration working example. Results are given for SVDGP and FlaGP using both MCMC and MAP estimation over a range of ensemble sizes and 10 random ensemble designs. In the top left, the interval score is shown. In the top right panel, the RMSE is shown. On the bottom left, the total time (in log seconds) required for model fitting and prediction.}
    \label{fig:calibration_results}
\end{figure}

\Cref{fig:calibration_results} shows the test set RMSE in the top right panel, which shows that the accuracy of calibrated predictions is nearly identical between FlaGP and SVDGP. Similarly, the interval score in the top left panel shows that both models give very similar uncertainty in their calibrated predictions. Where these models differ is in their computational requirements. In the bottom left panel, we see that the SVDGP model is more than an order of magnitude slower for even the smallest $M$ and about 2.5 orders of magnitude slower for $M=1000$. These results illustrate that calibration with SVDGP becomes quickly infeasible as the ensemble size increases, while FlaGP remains fast, requiring just over 100 seconds for $M=10000$.

\Cref{tbl:ub_coverage} shows that FlaGP and SVDGP gives near nominal coverage over the range of ensemble sizes, both tending to slightly under-cover.

\begin{table}
\begin{center}
\begin{tabular}{|c|c|c|c|c|c|c|c|}
\cline{3-8}
\multicolumn{2}{c|}{} & \multicolumn{6}{c|}{$M$} \\
\cline{3-8}
\multicolumn{2}{c|}{} & 100 & 250 & 500 & 1000 & 5000 & 10000 \\
\cline{3-8}
\hline
\multirow{2}{*}{$Model$} & FlaGP & $92\%$ & $94\%$ & $93\%$ & $94\%$ & $93\%$ & $93\%$  \\
\cline{2-8}
& SVDGP & $92\%$ & $94\%$ & $93\%$ & $94\%$ & &   \\
\hline
\end{tabular}
\caption{Posterior coverage probability for true calibration parameter $\theta$ over a range of ensemble sizes. For all ensembles, the same $n=5$ observations are used for calibration. For FlaGP, $m=20$ remains fixed.}
    \label{tbl:ub_coverage}
\end{center}
\end{table}

\subsection{Biased Calibration}\label{sec:biased_calibration}
Turning to the situation where the calibrated computer model is thought to be a biased estimate of the mean of the field data, the data model in \Cref{eqn:ub_model} is adapted in the following way,
%\begin{align}\label{eqn:func_y_dist_biased}
  %  \bm{y}(\bm{x}_i^F)\;|\;\bm{\eta}(\bm{x}_i^F,\bm{\theta}),\bm{\delta}(\bm{x}_i^F),\bm{\theta},\sigma^2_y\sim N(\bm{\eta}(\bm{x}_i^F,\bm{\theta})+\bm{\delta}(\bm{x}_i^F),\sigma^2_y\bm{I}_{d_y});\;i=1,\ldots,n,
%\end{align}

\begin{equation}\label{eqn:biased_model}
    \bm{y}(\bm{x}_i^F) = \bm{\eta}(\bm{x}_i^F,\bm{\theta}) + \bm{\delta}(\bm{x}_i^F) + \bm{\epsilon};\;\bm{\epsilon} \sim N(\bm{0},\sigma^2_y\bm{I}_{d_y}),
\end{equation}
where $\bm\delta(\bm{x})$ accounts for the systematic bias between the emulator and the field observations. 

The emulator $\bm{\eta}$ is the same as \Cref{eqn:em_form}, and similarly, the discrepancy is modeled as
\begin{equation}\label{eqn:delta_sum}
    \bm\delta(\bm{x}) = \sum_{j=1}^{p_{\delta}} \bm{k}_j v_j(\bm{x}),
\end{equation}
where $\{\bm{k}_1,\ldots,\bm{k}_{p_{\delta}}\}$ is a collection of $d_y$-dimensional basis vectors with scalar weights $v_j(\bm{x})$. Conditional on the emulator, the matrix of weights is calculated as $\bm{V}^T = (\bm{K}^T\bm{K})^{-1}\bm{K}^T\bm{R}_y$, where $\bm{R}_y$ is a matrix of residuals with columns $\bm{y}(\bm{x}_i^F)\;-\;\bm{\eta}(\bm{x}_i^F,\bm{\theta})$. By default, a full GP is used to model these weights rather than laGP, because $n$ is usually small enough that computation is not an issue. We use the full GP implementation in the \texttt{laGP} package, which uses a separable Gaussian correlation function and the same empirical Bayes procedure for lengthscale estimation discussed in Section \ref{sec:input_scaling}.

Basis vectors for the discrepancy model must be carefully selected both to achieve sensible discrepancy models and to avoid over-fitting the relatively small number of field data points available for training the discrepancy model. %One approach is to predict the observed data using the emulator and use the SVD of the residuals to select basis functions. We found that basis vectors selected in this way can result in severely over-fit discrepancy models, especially for small $n$, given the flexibility of a GP. 
For this work, we adopt the approach of \cite{higdon} by using user-selected basis functions based on prior knowledge of the discrepancy model form.

 %Scaled inputs are not used here %to bypass parameter estimation as the response values have changed, and are conditional on the emulator prediction.

Like in the unbiased calibration case, we approximate the posterior predictive distribution of the emulator and discrepancy model as Gaussian, giving the data likelihood
\begin{equation}\label{eqn:func_y_dist_biased}
\begin{aligned}
    %\bm{y}(\bm{x}_i^F)\;|\bm \eta(\bm x_i^F,\bm\theta),\bm{\delta}(\bm{x}_i^F),\bm{\theta},\sigma^2_y \sim N(&\bm\mu_{\bm{\eta}}(\bm{x}_i^F,\bm{\theta}) + \bm\mu_{\bm\delta}(\bm x_i^F), \bm{\Sigma}_{\bm{\eta}}(\bm{x}_i^F,\bm{\theta}) + \bm{\Sigma}_{\bm{\delta}}(\bm{x}_i^F) + \sigma^2_y\bm{I}_{d_y});\;i=1,\ldots,n,
    \bm{y}(\bm{x}_i^F)\;|\;\cdot \sim N(&\bm\mu_{\bm{\eta}}(\bm{x}_i^F,\bm{\theta}) + \bm\mu_{\bm\delta}(\bm x_i^F), \bm{\Sigma}_{\bm{\eta}}(\bm{x}_i^F,\bm{\theta}) + \bm{\Sigma}_{\bm{\delta}}(\bm{x}_i^F) + \sigma^2_y\bm{I}_{d_y});\;i=1,\ldots,n,
    \end{aligned}
\end{equation}
where $\bm{\Sigma}_{\bm{\eta}}(\bm{x}_i^F,\bm{\theta}) = \bm B^T\bm \Sigma_{\bm w}\bm{B}$ and $\bm{\Sigma}_{\bm{\delta}}(\bm{x}_i^F) = \bm K^T\bm \Sigma_{\bm v}(\bm{x}_i^F)\bm K$. Here, $\bm \Sigma_{\bm v}(\bm{x}_i^F)$ is the predictive covariance matrix for the discrepancy model at the observed inputs $\bm{x}_i^F$, consisting.

%Following similar notation as in \Cref{eqn:func_y_dist}, $\bm\mu_{\delta}(\bm x_i^F)$, and $\bm\Sigma_{\delta}(\bm x_i^F)$ are the posterior predictive mean and covariance of the discrepancy model at input $\bm x_i^F$. 

Analogous to \Cref{eqn:func_post}, the posterior conditional density of $\bm{\theta}$ and $\sigma^2_y$ is then
\begin{equation}\label{eqn:func_post_biased}
    p(\bm{\theta},\sigma^2_y|\cdot) \propto \prod_{i=1}^n p(\bm{y}(\bm{x}_i^F)\;|\;\cdot) \times p(\bm{\theta}) \times p(\sigma^2_y),
\end{equation}
where %$p(\bm{y}(\bm{x}_i)|\bm{\eta}(\bm x_i,\bm\theta),\bm{\delta}(\bm x_i),\bm{\theta},\sigma^2_y)$
$p(\bm{y}(\bm{x}_i)\;|\;\cdot)$ is the likelihood computed from the conditional normal distribution in \Cref{eqn:func_y_dist_biased} and $p(\bm{\theta})$, $p(\sigma^2_y)$ are the same prior distributions defined in Section \ref{sec:unbiased_calibration}. The following algorithm is used to sample from $p(\bm{\theta},\sigma^2_y|\cdot)$ , assuming a matrix $\bm{K}=[\bm{k}_1|\dots|\bm{k}_{p_{\delta}}]$ of basis functions for the discrepancy model has been predefined. A more detailed overview of the calibration algorithm is given in Section SM1 of the supplementary material.

%compute the logarithm of the RHS of \Cref{eqn:func_post_biased} for a given $\bm{\theta}$ and $\sigma^2_y$. %which is necessary for inference on $\bm{\theta}$. 
\begin{algorithm}
\caption{Calibration with a discrepancy model}\label{alg:biased_calib}
\begin{algorithmic}[1]
\itemsep0em
\item \textit{Precompute FlaGP Emulator}: Steps $1,2,$ and $3$ of \Cref{alg:em_form}
\item \textit{Sample parameters}: Given a current value of the parameters, $\bm{\psi}_{\text{curr}} = [logit(\bm{\theta}),log(\sigma^2_y)]$, sample a new value $\bm{\psi}_{\text{new}}$. Since $\bm{\theta}\in[0,1]^{d_x}$ and $\sigma^2_y>0$, we use a multivariate normal distribution to propose jointly $\bm{\psi}_{\text{new}} \sim \text{N}(\bm{\psi}_{\text{curr}},\bm{\Sigma}_{\text{prop}})$
\item \textit{Predict Observations}: Use the scaled nearest neighbor strategy in step 4 of \Cref{alg:em_form}, compute $\bm{\mu}_{\bm{\eta}}(\bm{x}_i^F,\bm{\theta})$ and $\bm{\Sigma_{\bm{\eta}}}(\bm{x}_i^F,\bm{\theta})$ for $i=1,\ldots,n$.
\item \textit{Residuals}: Compute $\bm{y}(\bm{x}^F_i)-\bm{\mu}_{\bm{\eta}}(\bm{x}_i^F,\bm{\theta});\;i=1,\ldots,n$ and build residual matrix $\bm{R}_y$.
\item \textit{Discrepancy precomputing}: Compute discrepancy basis weights $\bm{V}^T = (\bm{K}^T\bm{K})^{-1}\bm{K}^T\bm{R}_y$. %Then $\bm{R}_y=\bm{K}\bm{V}^T$.
\item \textit{Discrepancy prediction}: Fit $p_\delta$ independent separable GP models to the rows of $\bm{V}^T$ and compute predictive mean $\bm{\mu}_{\bm{\delta}}(\bm{x}_i^F)$ and covariance matrix $\bm{\Sigma}_{\bm{\delta}}(\bm{x}_i^F)$ at observed input locations $\bm{x}_1^F,\ldots,\bm{x}_n^F$.
%\item \textit{Compute likelihood}: With new residuals $\bm{y}(\bm{x}^F_i) - \bm{\eta}(\bm{x}^F_i,\bm{\theta}) - \bm{\delta}(\bm{x}^F_i)$ compute the conditional normal likelihood using Eq. \eqref{eqn:func_y_dist_biased}.
\item \textit{Posterior density}: Compute the log posterior density (\Cref{eqn:func_post_biased}) using the conditional normal likelihood in \Cref{eqn:func_y_dist_biased}.
\item \textit{Repeat sampling}: Repeat steps $2-4$ in a Metropolis-Hastings MCMC (MH-MCMC) sampler for posterior inference on $\bm{\theta}$ and $\sigma^2_y$.
\end{algorithmic}
\end{algorithm}

\subsubsection{Working Example}\label{sec:biased_working_ex}

%Our working example is used to illustrate biased calibration. With a slight modification to \Cref{eqn:ub_datagen}, the computer model becomes a biased representation of the mean of the simulated field observations. In this example, computer model output is generated from
%\begin{equation}\label{eqn:bias_datagen}
%    f(x,t,\tau) = \frac{t\text{sin}(xt\tau)}{3\tau} + (\tau-1)^3
%\end{equation}
%while the observed data are still generated from \Cref{eqn:ub_datagen}. 

We return to the working example to illustrate calibration with discrepancy. Biased observations are generated directly from the unbiased observations in \Cref{sec:unbiased_working_ex}. For functional data, defining basis vectors for a discrepancy model is non-trivial, requiring either strong prior knowledge or the use of an extremely flexible basis able to capture a range of discrepancies. A flexible basis is often ill-advised unless many field observations are available for informing the discrepancy model. When only a small number of observations are available (in this case, $n=5$), we generally recommend defining a simple basis using prior knowledge. For this illustrative example, we apply a simple linear discrepancy to the unbiased observations and specify linear basis vectors ($p_\delta=2$) for the discrepancy model. Examples of these biased observations are seen in blue in \Cref{fig:ensemble_vis}.

\begin{figure}
\centering
%\begin{subfigure}{0.35\textwidth}
%  \includegraphics[width=\linewidth]{images/lagp_mcmc.pdf}
%  \caption{FlaGP}
%  \label{fig:bias_mcmc_lagp}
%\end{subfigure}%
%\hspace*{\fill}
%\begin{subfigure}{0.35\textwidth}
%  \includegraphics[width=\linewidth]{images/sepia_mcmc.pdf}
%  \caption{SVDGP}
%  \label{fig:bias_mcmc_sepia}
%\end{subfigure}
\includegraphics[width=.75\linewidth]{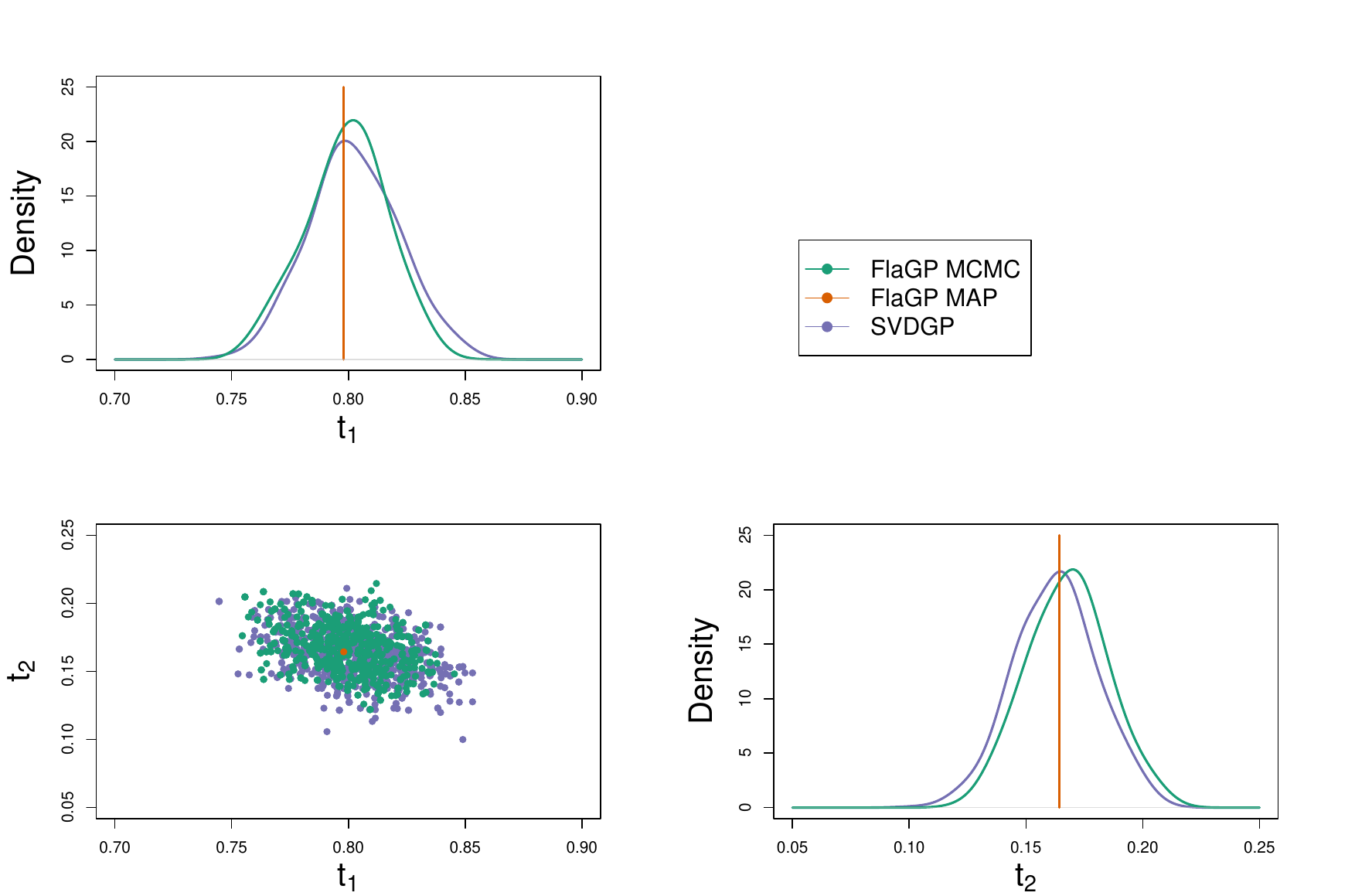}
\caption{\small Marginal posterior distributions and bivariate posterior MCMC samples for the biased calibration working example with $M=100$. MAP estimates from FlaGP are shown. MAP estimation is discussed in \Cref{sec:fast_pt_est}.}
\label{fig:biased_mcmc}
\end{figure}

Note that we do not expect either model to recover the true parameter values due to the discrepancy model \citep{loeppky2006computer,calib_bias_tuo}. \Cref{fig:biased_mcmc} shows posterior density estimates for a single ensemble design. Since FlaGP uses a modular emulator, and SVDGP fits the emulator and discrepancy models jointly, we are not surprised that posterior distributions for the two models are slightly different. 

\begin{figure}
\centering
\includegraphics[width=\linewidth]{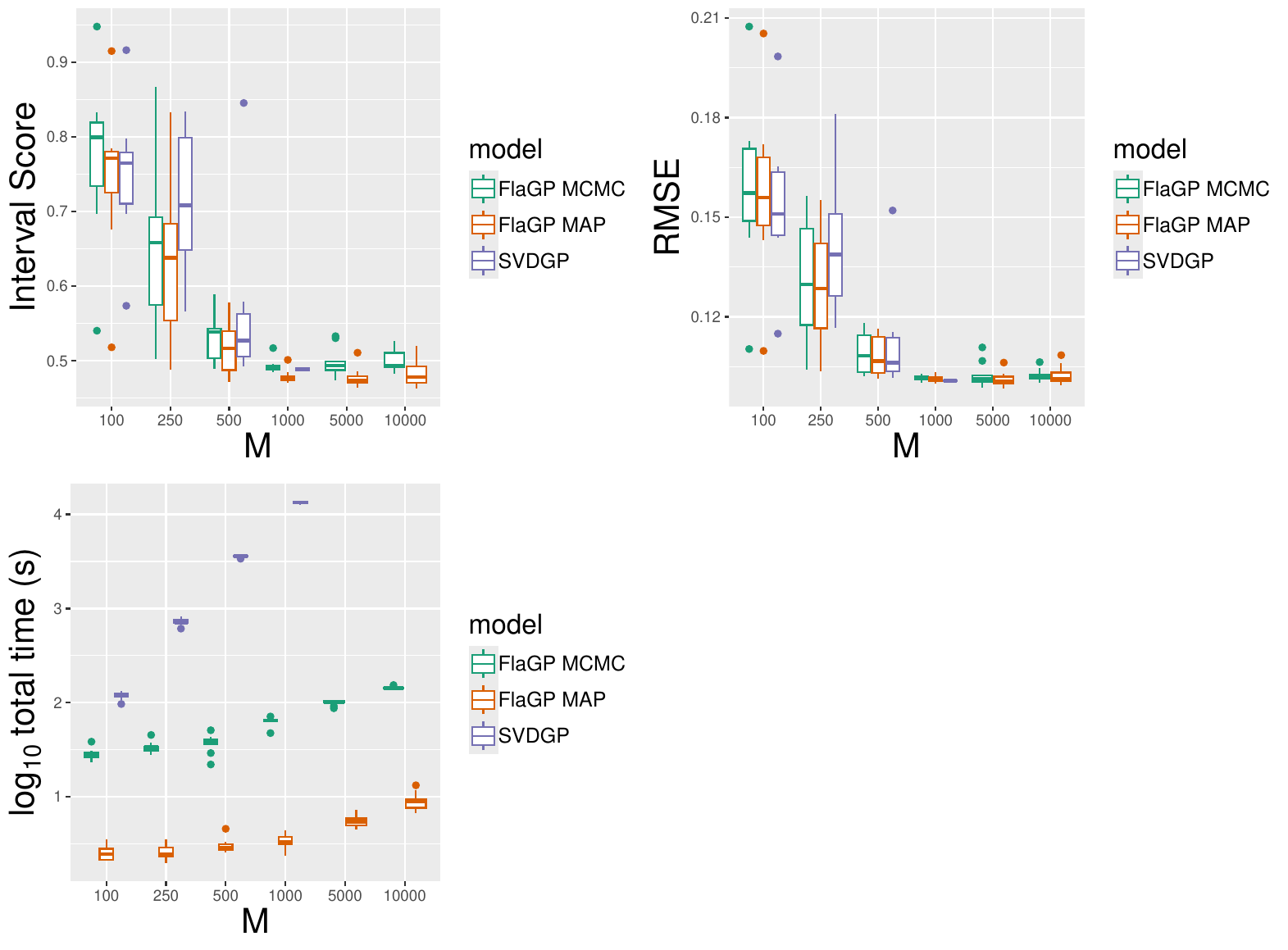}
\caption{\small Biased calibration working example posterior predictive scoring and computation time.}
\label{fig:biased_scores}
\end{figure}

For each run of the study, predictive performance is evaluated on a test set of 100 curves generated from \Cref{eq:working_example}. Test set predictive scores are presented in Fig.~\ref{fig:biased_scores}, averaged over the 100 test curves. FlaGP and SVDGP perform very similarly for both RMSE and IS with the exception of SVDGP slightly under-performing at $M=250$. For the other ensemble sizes where SVDGP was run, it may give very slight improvements in RMSE, but these differences are unlikely to be significant given the increased computational cost.

For $M=1000$, FlaGP requires about 65 seconds for precomputing, 5000 posterior samples using $m_c=20$, and 100 posterior predictive samples for each of the 100 test curves using $m=50$. SVDGP, on the other hand, requires over 3.7 hours. The predictive scores in \Cref{fig:biased_scores} indicate that for these data, the potential improvements from SVDGP are minimal, or non-existent. For more complicated problems, we may see some improvements with SVDGP, but our experience is that those improvements are small and are outweighed by the reduction in computational cost for large ensemble datasets.

\subsection{Fast Point Estimation}\label{sec:fast_pt_est}

The proposed methodology is designed to facilitate very fast MCMC sampling so that uncertainty in the calibration parameters can be quickly estimated. However, there are problems for which a fast MAP estimate of the calibration parameters may be sufficient. Furthermore, in cases where the posterior distribution is approximately multivariate normal, a good approximation to the posterior can be obtained rapidly via a Laplace approximation \citep{laplace_approx}, although we have found robust estimation of the Laplace approximation to be challenging when $p_t>10$.

To compute a MAP estimate of the calibration parameters, \Cref{eqn:func_post} or \Cref{eqn:func_post_biased} is optimized with respect to $\bm\theta,\sigma_y^2$. Fast likelihood evaluation is important, and as discussed in \Cref{sec:unbiased_calibration}, an application of $\textit{Result 1.}$ from \cite{higdon}, allows for likelihood evaluation in $\mathcal O((n(p_\eta+p_\delta))^3)$ operations. Likelihood evaluation tends to be very fast for most problems we encounter, and will only become a computational issue when either a large number of experiments are available, or many basis components are needed to capture the simulation outputs. %\textcolor{blue}{If $n(p_\eta+p_\delta)>1000$, there are other options for speeding up likelihood calculations, which are discussed in Section SM4 of the supplementary material.}

An important requirement to apply $\textit{Result 1.}$ is that the posterior predictive distribution from the emulator (or emulator + discrepancy model) follows a Normal distribution. Our framework uses laGP for fast emulation, which results in a Student-t predictive distribution due to integrating out the process variance. To apply $\textit{Result 1.}$, we approximate the Student-t predictive distributions with a Normal distribution (see \Cref{sec:eff_llh_calc}). %The details of the derivation can be found in \Cref{sec:eff_llh_calc}.

The Normal approximation to the Student-t will lead to a negligible reduction in uncertainty when the number of neighbors used in prediction is not extremely small. Given that the methods are designed for large ensembles, it is reasonable to expect that the degrees of freedom for the Student-t predictive distribution will be large enough that the Normal approximation will be very good ($m>20$).

%It is important to note that the approximation can cause underestimation of the predictive variance for the discrepancy model, especially for small $n$. This is compensated for by slightly larger estimates of $\sigma^2_y$. By using these sampling distributions to redefine the conditional likelihoods $p(\bm{y}(\bm{x}_i)|\cdot)$, posterior estimation can be efficiently carried out. In the unbiased scenario, this is achieved by letting Eq. \eqref{eqn:func_post} be the objective function an the optimization algorithm. 

For this work, the NOMAD \citep{NOMAD} algorithm is used for optimization. We found that NOMAD strikes a good balance between computation and robust optimization, especially for $p_t>10$. For small $p_t$, we tend to use \texttt{optim} in \texttt{R}, which provides a Laplace approximation to the posterior distribution. %Similarly, in the biased scenario, Eq. \eqref{eqn:func_post_biased} is employed for the same purpose.

%\vspace{-4mm}
\subsubsection{Working Example}\label{sec:map_ball_drop}

In the unbiased calibration example (\Cref{sec:unbiased_working_ex}), \Cref{fig:calibration_results} shows that (i) the predictive mean and uncertainty quantification from the MAP approach is on par with the other two methods, and (ii) the MAP approach is notably faster and can scale well with $M$.

For the biased calibration example (\Cref{sec:biased_working_ex}), \Cref{fig:biased_scores} again shows that the MAP approach is very competitive but and may provide improved prediction UQ. While these results are encouraging for fast emulation and calibration without MCMC, there remains the drawback of no uncertainty in the estimates of the calibration parameters.%, which calls the overall UQ into question. Furthermore, this approach cannot give insight or quantify uncertainty associated with multimodal posterior distributions. 
If UQ in the calibration parameters is not needed, the computational benefits from the MAP approach cannot be ignored, being an order of magnitude faster than the MCMC approach.

%\vspace{-4mm}
\section{Al-5083 Analysis}\label{sec:Al_example}
%\vspace{-4mm}

The FlaGP methodology is applied to the application dataset described in Section \ref{sec:Application}. Recall that an ensemble of $M=20000$ runs from the FLAG simulator is available for three experimental setups, which vary the flyer-plate and sample thickness. The output of each simulation is a velocity-time curve. A single experimental velocity curve is also available for each shot. As in \cite{higdon_flyer_plate}, the three shots are emulated jointly, resulting in a functional response of length $d_y=600$. Separate emulators could be used for each experimental setup, with a product likelihood over the emulators used for calibration; however, we find that emulation is only slightly improved for shot 104, and is worse for shots 105 and 106 under that framework, indicating that sharing information between the three shots is beneficial.

The ensemble is sufficiently large so that the SVDGP method is computationally infeasible to apply; however, a tractable analysis can be done with the proposed method. Recall that a calibration study for similar data was done in \cite{higdon_flyer_plate} with SVDGP; however, they reduced the velocity-time curves to $d_y=12$ to make the analysis feasible, and we estimate that it may have taken over 10 days on a high-performance computer. This paper addresses the scenario where emulation and calibration must be fast, and analysis can be carried out on a personal computer (e.g. a laptop or desktop) rather than requiring high-performance computing resources. As such, the analyses presented in this section are all done on an Apple M2 Max MacBook Pro.

In Section \ref{sec:al_20k_em}, we show that the proposed emulator can quickly and accurately emulate FLAG, in \Cref{sec:al_active_learning} we show that the active learning framework developed in \Cref{sec:seq_d} could be used to improve emulation for smaller ensemble sizes, and in Section \ref{sec:al_20k_calib}, we present a fast calibration that accurately predicts the observed data. This application is representative of the type used in production at materials science facilities. For these applications, the FlaGP methodology provides calibration that can be updated very quickly in the presence of additional experimental data.

%The FlaGP methodology is applied to the application dataset described in Section \ref{sec:Application}. As in \cite{higdon_flyer_plate}, the 3 shots are used jointly for analysis giving $d_y=1500$. %Emulating the densely sampled curve, rather than a small feature set adds the complication of near zero values at early time indices. To account for this and prevent negative predictions, we model the square root of the velocity, and transform predictions back to the original scale. 
%This ensemble is sufficiently large so that the SVDGP method is computationally infeasible to apply, however a tractable analysis can be done with the proposed method. In Section \ref{sec:al_20k_em} we show that the proposed emulator can quickly and accurately emulate FLAG, and in Section \ref{sec:al_20k_calib} we present a fast calibration that accurately predicts the observed data.
%\vspace{-4mm}
\subsection{Emulation}\label{sec:al_20k_em}
To assess emulator accuracy, the FlaGP emulator is fit to 19000 simulations, and 1000 are kept as a prediction hold-out set. To reduce the possibility of extrapolation, the test set is randomly selected from within the 2.5\% and 97.5\% quantiles of the marginal input distributions. 

The first step in fitting the emulator is to obtain the empirical basis functions using the $SVD$ of the computer model output. Given the large ensemble data matrix, a random SVD (RSVD) \citep{random_svd} could be used rather than a full SVD. In this case, $M\times d_y$ is small enough that a full SVD only requires about 8 seconds. If the ensemble were larger or the functions sampled at more points, the RSVD would be more useful. We considered the RSVD and found that emulation and calibration were not noticeably affected. For the analysis, 8 basis vectors, capturing $99\%$ of the variability in the responses, were used.

%The RSVD approximates the first 7 basis vectors in less than 3 seconds, 20 times faster than a full SVD. This number of basis vectors was chosen so that over $99\%$ of the variability in the data is captured. For these data the RSVD basis vectors are an extremely accurate representation of the full SVD basis. %A full SVD takes about 20 times longer for these data and we found that the RSVD basis vectors are an extremely representation of the full SVD basis. 

Given a basis decomposition, the next step is to scale the input space for fast prediction. Stratified random sampling is used to select subsets of the ensemble for lengthscale estimation.%, as BLHS can struggle for large $d_x$. 
Parameters $m_\text{est}$ and $r_\text{est}$ must be chosen, and a reasonable approach is to select small initial values, resulting in fast estimation. Parameter values can then be increased if predictive metrics indicate that improvement is needed. Given the exploratory nature of this example, we have the computational budget to explore a range of $m_\text{est}$ values ranging from $128$ to $1024$. %For these data we initially use $m_\text{est} = 256$. Since $m_\text{est}\ll M$, we select a fairly large $r_\text{est}=25$. With the RSVD, the total fitting time for this model is just over 1 minute.

To explore emulator accuracy on the hold-out set, we consider shot-specific RMSE averaged over the velocity curve for a range of $m_\text{est}$ and $m$. These are shown in \Cref{tbl:19k_em}. %, computed by dividing the RMSE by the average velocity of each shot. N
RMSE is used as the predictive scoring metric rather than relative error because there are many velocity values at or near zero, which present issues for relative error calculations. %Normalized RMSE was also considered, but it lacks the same level of interpretability. 

We see that RMSE tends to be the range of 4-9 m/s, which is acceptable given that the velocity curves range from $0-500$ m/s. The fact that the RMSE for shots \textit{105} and \textit{106} is larger than for shot \textit{104} does not indicate that they are poorly emulated by comparison. If we take the RMSE values that are bold in the table $4.37,8.04,$ and $8.78$, and divide by the range of the data for each shot we get normalized RMSE values of $0.0202, 0.0214, 0.0171$ indicating that all shots are emulated with very similar accuracy.%, and \textit{106} is emulated the most accurately.

%In fact, normalized RMSEs indicate that shot \textit{106} is the most accurately predicted. Dividing the RMSE values of 3.23, 4.88, and 5.26 by mean shot velocity gives relative RMSEs of 0.023, 0.025, and 0.018. %These results in Table \ref{tbl:19k_em} indicate that the FLAG simulator can be accurately emulated using the proposed approach in near real time.

\begin{table}
\begin{center}
\scalebox{0.85}{\begin{tabular}{|c|c|c|c|c|c|c|c|c|c|c|c|c|c|c|}
\cline{3-14}
\multicolumn{2}{c|}{} & \multicolumn{12}{c|}{$m_\text{est}$, \textit{shot}\#} \\
\cline{3-14}
\multicolumn{2}{c|}{} & 
\multicolumn{3}{c|}{128} & \multicolumn{3}{c|}{256} & \multicolumn{3}{c|}{512} & \multicolumn{3}{c|}{1024} \\
\cline{3-15}
\multicolumn{2}{c|}{} & 
\textit{104} & \textit{105} & \textit{106} & 
\textit{104} & \textit{105} & \textit{106} & 
\textit{104} & \textit{105} & \textit{106} & 
\textit{104} & \textit{105} & \textit{106} & Time (s) \\
\cline{3-15}
\hline
\multirow{5}{*}{$m$} 
& 25 & 
6.26 & 9.69 & 10.32& 
5.06& 8.52&  9.46&
4.71& 8.27 & 9.14&
4.77 &8.34 & 9.17 & 0.00132\\ % time is devided by 1000 (time per prediction)
\cline{2-15}
& 50 & 
5.85 &9.48&10.13&
4.82 &8.32 & 9.31&
\textbf{4.37} & \textbf{8.04} & \textbf{8.78}&
4.43& 7.93 & 9.18 & 0.00182\\
\cline{2-15}
& 100 & 
5.54 &9.11 &9.95&
5.05 &8.38 &9.32&
4.67& 8.11& 9.00&
4.54& 7.87 &9.37 & 0.00383\\
\cline{2-15}
& 250 & 
4.83& 8.22& 9.47&
4.59& 7.91& 9.11&
4.74& 8.17& 9.02&
4.31 &8.24 &9.20 & 0.02373\\
\cline{2-15}
& 500 &
4.84 &8.41& 9.23&
4.63& 8.07& 9.14&
4.73& 8.24 &9.19&
4.45& 8.14& 9.16 & 0.16524 \\
\hline
%\multicolumn{2}{c|}{Time (s)} & \multicolumn{3}{c|}{9.53} & \multicolumn{3}{c|}{66.10} & \multicolumn{3}{c|}{844.33} & \multicolumn{3}{c|}{11432.14} & \\ % total fitting time
\multicolumn{2}{c|}{Time (s)} & \multicolumn{3}{c|}{0.381} & \multicolumn{3}{c|}{2.644} & \multicolumn{3}{c|}{33.773} & \multicolumn{3}{c|}{457.286} & \\ % fitting time per bootstrap replicate
\hline
\end{tabular}}
\caption{Average RMSE (m/s) over 1000 hold-out velocity profiles for each of the three shots. All combinations of  $m_\text{est}\in\{128,256,512,1024\},m\in\{25,50,100,250,500\}$ are shown. $m_\text{est}=512$ with $m=50$ provides a good mix of accuracy and computational efficiency. The rightmost column gives average prediction times (per prediction) for a given $m$, and the bottommost row gives emulator fitting time per replicate under different $m_\text{est}$.} 
  \label{tbl:19k_em} 
\end{center}
\end{table}
\begin{comment}
\begin{table}[!h]
\begin{center}
\scalebox{0.7}{\begin{tabular}{|c|c|c|c|c|c|c|c|c|c|c|}
\cline{3-11}
\multicolumn{2}{c|}{} & \multicolumn{9}{c|}{$m_\text{est}$, \textit{shot}\#} \\
\cline{3-11}
\multicolumn{2}{c|}{} & \multicolumn{3}{c|}{256} & \multicolumn{3}{c|}{512} & \multicolumn{3}{c|}{1024} \\
\cline{3-11}
\multicolumn{2}{c|}{} & \textit{104} & \textit{105} & \textit{106} & \textit{104} & \textit{105} & \textit{106} & \textit{104} & \textit{105} & \textit{106} \\
\cline{3-11}
\hline
\multirow{3}{*}{$m$} & 50 & 3.23 & 4.88 & 5.26 & 3.11 & 4.60 & 5.11 & 3.10 & 4.70 & 5.02 \\
\cline{2-11}
& 100 & 3.16 & 4.80 & 5.19 & 3.14 & 4.50 & 5.08 & 3.21 & 4.70 & 5.07 \\
\cline{2-11}
& 250 & 3.04 & 4.44 & 4.91 & 3.00 & 4.26 & 4.95 & 3.01 & 4.35 & 4.93 \\
\hline
\end{tabular}}
\caption{Average RMSE $(m/s)$ over 1000 hold-out velocity profiles for each of the 3 shots. All combinations of  $m_\text{est}\in\{256,512,1024\},m\in\{50,100,250\}$ are shown.} 
  \label{tbl:19k_em} 
\end{center}
\end{table}
\end{comment}

Table \ref{tbl:19k_em} shows that accurate emulation can be achieved with a very small computational budget. Larger computational budgets are considered with increased $m_\text{est}$ and $m$, but the improvements in predictive accuracy may not outweigh the increased computational cost. The results show that there is some gain in performance up to a point, but there may be no practical significance, given the increased computational cost. It is also important to notice that accuracy quickly plateaus. For example, $m_\text{est}=512$, about 34 seconds are needed for the precomputing phase for each estimation resample, and for $m_\text{est}=1024$, this increases to nearly 8 minutes. For prediction, $m=250$ is more than an order of magnitude slower than $m=50$ and provides no improvement. The results in Table \ref{tbl:19k_em} indicate that the FLAG simulator can be accurately emulated with FlaGP given a very small computational budget, and only small improvements are achieved with an increased budget. In \Cref{sec:seq_d_working_ex} we briefly discussed the approximate computational cost of active learning for these data, but we do not explore it for emulator improvement as the computational budget for additional runs of the FLAG simulator is not available.

\subsection{Active Learning}\label{sec:al_active_learning}

Here we explore the effect of active learning for the FLAG simulator. The ensemble of runs for this example was already defined using an LHS, and active learning was not considered. We present this as an exercise to explore the methods developed in \Cref{sec:seq_d}, and the trends found for the working example in \Cref{sec:seq_d_working_ex}. In \Cref{fig:sequential_design} we saw that sequential maximin designs performed as well as the more targeted IMSPE-based approach at a fraction of the cost. We hypothesized that this was be due to the simplicity of the emulation task. Here, we tackle the far more challenging task of emulating the FLAG simulator with $d_x=11$ inputs and $p_\eta=8$ basis components.

Our procedure is as follows. First, we take a random subset of 6110 runs from the full 20000 run design, which is done purely for computational efficiency in selecting the remaining data subsets. We take $M_\text{init}=110$ runs for training the initial emulator, which are selected using a sequential maximin procedure, initialized at the most central point of the 6110 run subset. With the remaining 6000 runs, 3 subsets of size 2000 are selected, again using sequential maximin initialized at a central point to ensure that the subsets are space filling. These 3 subsets are used for integration, candidate, and testing points. The integration points are used to evaluate the IMSPE criterion, the candidate points are potential points to be added to the training set, and the test points are used to determine the effectiveness of the active learning procedure through the models predictive RMSE.

Starting with such a small set of initial points in 11 dimensions leads to rapid and significant improvement in model performance with active learning. In total, we add 500 new points from the candidate set using the IMSPE, maximin, and scaled maximin criteria described in \Cref{sec:seq_d}. We chose to refit the emulator every time 50 points are added to the training set.

\begin{figure}
	\centering
	\includegraphics[width=\linewidth]{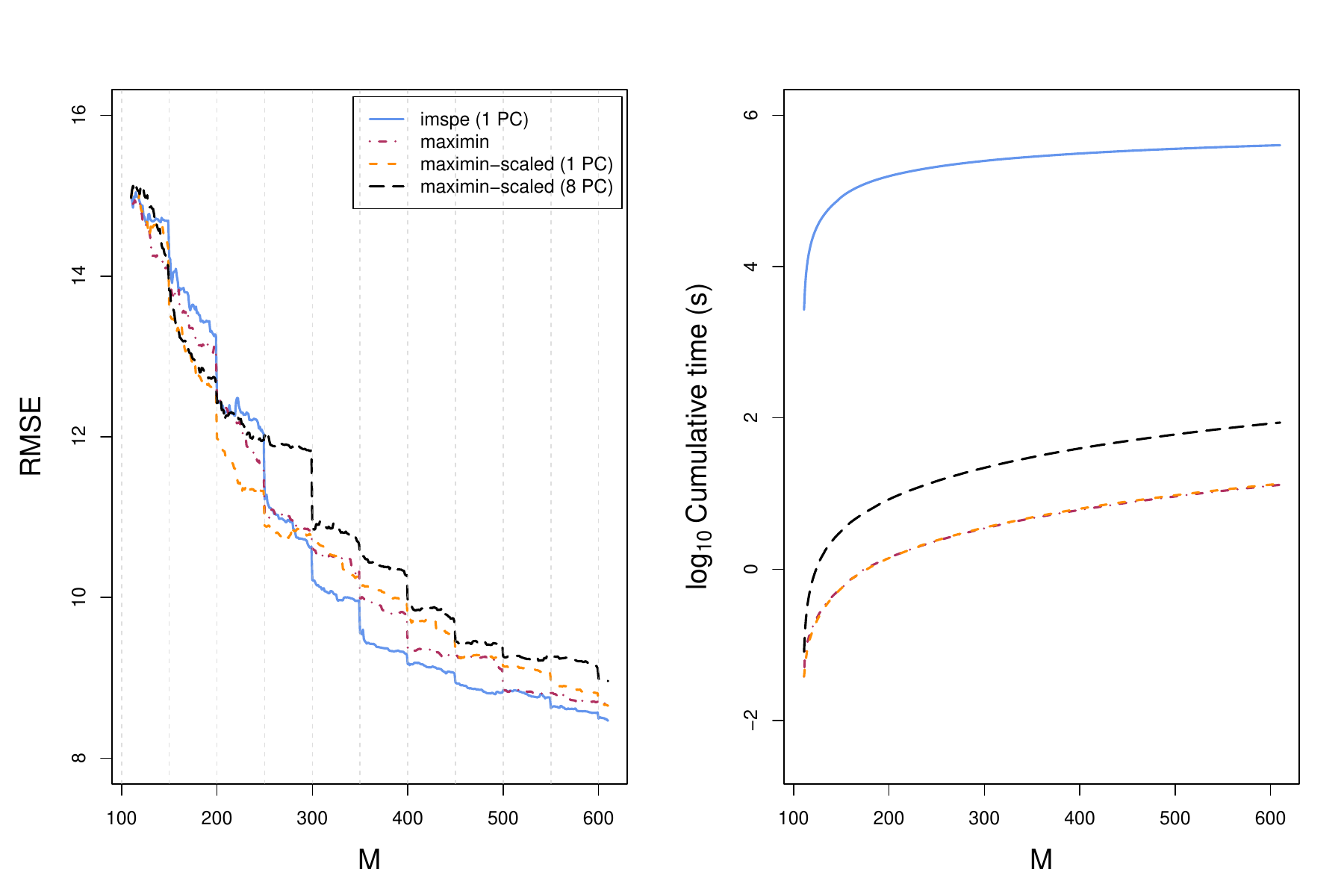}
	\caption{Active learning for FLAG simulator beginning at $M=110$, and adding 500 points sequentially. IMSPE appears to provide a benefit over the maximin approaches, but is computationally intensive by comparison. Vertical dashed lines indicate where model refitting takes place, often leading to a substantial drop in RMSE.}
	\label{fig:Al_active_learning}
\end{figure}

\Cref{fig:Al_active_learning} shows test set RMSE on the left and cumulative computation time on the right. Unlike the working example in \Cref{sec:seq_d_working_ex}, IMSPE-based active learning gives some benefit over the faster maximin approaches. We still do not see any notable benefit from leveraging input scaling for the maximin approach, however. While IMSPE may give better designs, the difference in computation time may not make it practically worth the effort. In this case, the cost of adding the first point using the IMSPE criterion with 1 basis component is .75 hours, and the total cost of adding 500 points is nearly 5 days. Given that, we did not explore a more expensive version with additional components, nor did we consider randomizing this procedure over different data subsets. On the other hand, all maximin based approaches are extremely fast, ranging from about .01 to .22 seconds per point. 

It is important to consider that the sequential nature of these algorithms, and stochasticity related to model refitting, makes comparison difficult. In order to say with more certainty that the IMSPE criterion provides notable improvement over maximin, the algorithm should be initialized with many different subsets of the data. Such an in depth comparison is beyond the scope of this work, but is an important direction for future studies that would demand significant practitioner effort and computational resources.

The result in \Cref{fig:Al_active_learning} indicate that an active learning approach to design may provide notable emulator improvement for the same number of runs of the FLAG simulator. Consider that the IMSPE-based design achieves a test set RMSE of about 8.5 at $M=600$, and appears to be quickly approaching the performance of the emulator based on the full ensemble (see \Cref{tbl:19k_em}). Those results did use a different test set, so they are not directly comparable, but the results are promising nonetheless.

\subsection{Calibration}\label{sec:al_20k_calib}

In this section, we use the emulation methods described above, along with the experimental velocity curves, to constrain the 11 input parameters. Section \ref{sec:al_20k_em} informs reasonable choices of tuning parameters $m_\text{est}=512$, $r_\text{est}=25$, and $m=50$. Again, 8 basis vectors are used, computed with the SVD, and stratified random sampling is used to select subsets for lengthscale estimation. 

\begin{figure}
  \centerline{\includegraphics[scale=.6]{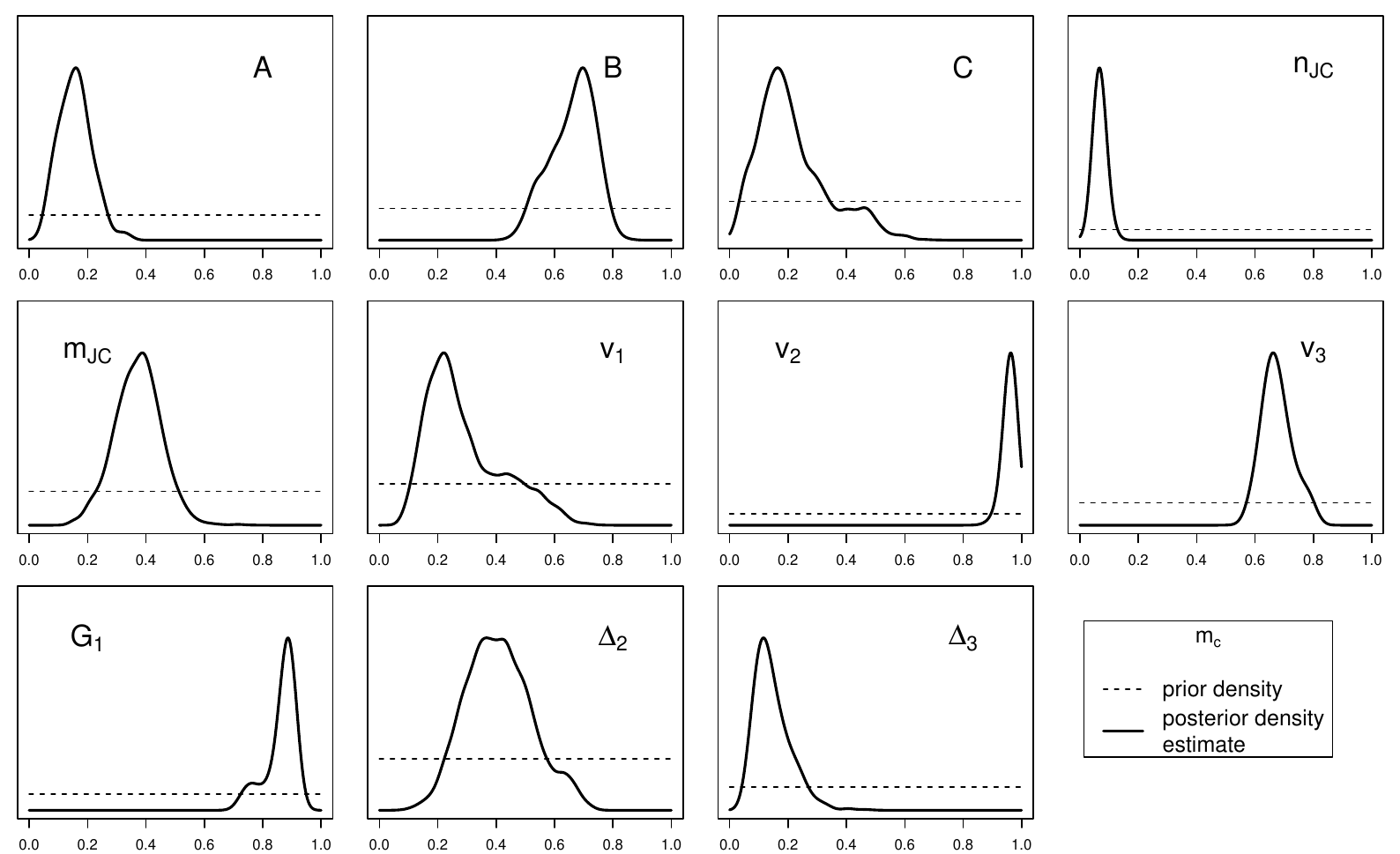}}
  \caption{Posterior density estimates of the calibration parameters for $m_c=50$.}
  \label{fig:Al_post_dens}
  %\vspace{-4mm}
\end{figure}

A modest budget is given for MCMC calibration, collecting 50000 samples, and using the first 25000 for proposal covariance adaptation. The total cost of precomputing and MCMC calibration is $14+52=66$ minutes. Uniform prior densities and posterior density estimates from the $25000$ post-burn-in samples are shown in Fig.~\ref{fig:Al_post_dens}. There is a notable reduction in prior uncertainty for all parameters. A very fast MAP analysis is also considered, which requires only about 3 minutes (with 10 random restarts of the optimizer) given the fitted emulator, giving a total cost of $14+3=17$ minutes.

%We also consider an analysis where the computational budget is severely constrained, letting $m_\text{est}=256$, $r_{est}=10$, and $m_c=50$.

%This analysis uses conservative choices of $m_\text{est}$, $r_\text{est}$, and the number of MCMC samples to ensure model convergence and accurate lengthscale estimation. A significantly more efficient analysis is done with $m_\text{est}=256$, $r_\text{est}=10$, and $m_c=50$, reducing precomputing time to less than 1 minute. By collecting only 10000 MCMC samples (using the first 5000 for covariance adaption), the total time is reduced to just 11.5 minutes. 

%A fast MAP estimate of the calibration parameters is also considered, which takes only 90 seconds with 10 random restarts of the optimizer, selected to match available parallel compute resources.

\begin{figure}
  \centerline{\includegraphics[scale=.6]{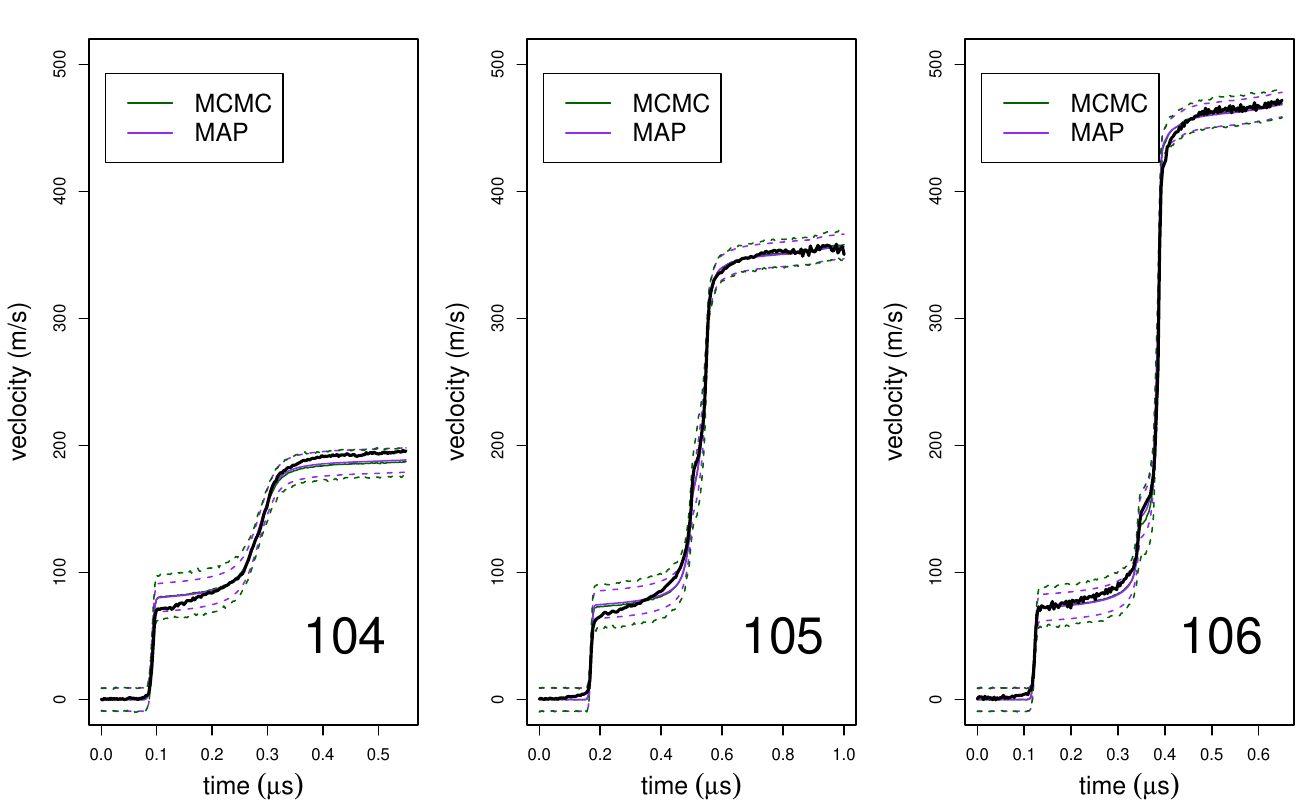}}
  \caption{Posterior predictive mean and 95\% prediction intervals for the observed shots using 1000 posterior MCMC samples or MAP estimate. The emulator is fit and calibrated using $m_\text{est}=512$, $m_c=50$, and predictions use $m=50$.}%using 3 modeling procedures. The top row shows predictions from an MCMC calibration with a moderate budget (2 hours), predictions from a faster MCMC calibration (12 minutes) are shown in row 2, and MAP calibration predictions in the bottom row (3 minutes).}
  \label{fig:al_big_calib_pred_obs}
  %\vspace{-4mm}
\end{figure}

Calibration is assessed on calibrated predictions, that is, how well calibrated emulators predict the observed data. \Cref{fig:al_big_calib_pred_obs} shows posterior predictive means and $95\%$ prediction intervals for both the MCMC and MAP analysis. Visually, the posterior means from both methods are very similar, but the MAP approach gives slightly narrower prediction intervals due to the lack of parameter uncertainty. Both models capture the observed data well, but are conservative in their estimate of uncertainty, giving greater than nominal coverage (see \Cref{tbl:calib_pred_obs}), likely due to model form error. This is most clearly seen in the prediction of shots 104 and 105, where the observations increase more steadily after the first velocity spike, while the predictions are flatter. We choose not to fit a discrepancy model because there is only a single observation for each shot. %Furthermore, without strong, physically motivated prior information, a discrepancy model would reduce the physical interpretation of the estimated parameters, which is not in line with the scientific aims of the application.

%A discrepancy model can certainly provide improved prediction; however, the shape of the true discrepancy is likely complicated here, and a discrepancy estimated from a single experiment is not likely to be generalizable. Furthermore, }

In \Cref{tbl:calib_pred_obs}, the RMSE, 95\% empirical coverage, and median IS are shown. The MAP calibration method gives improved prediction for shots 104 and 105, and its narrower intervals lead to improved interval score and coverage. Interval score is difficult to interpret on an absolute scale, and is best viewed as a relative comparison between models. For these data and modeling choices, the MCMC and MAP methods give very similar predictions. Choice between them may be determined based on computation and whether posterior distributions for the calibration parameters are needed. %s conservative uncertainties, and the MAP method may be preferable given its faster calibration time.

What is notable about both methods is that accurate, calibrated predictions are achieved in a relatively short amount of time with very modest computing resources (a personal laptop). In the pretense of new observations, posterior distributions can be updated very quickly.

%The first row of Fig.~\ref{fig:al_big_calib_pred_obs} shows posterior predictive means and $95\%$ intervals for the 3 observed velocity profiles using the moderate budget analysis with $m=50$ for prediction. All 3 shots are well predicted, and are mostly contained within the prediction intervals. The second and third rows show predictions for the faster MCMC and MAP models. Visually, there is very little difference between these predictions. The RMSE, IS, and computing time for all three models are given in Table \ref{tbl:calib_pred_obs}. The slower model does not score significantly better than the more efficient models, which was also seen in Section \ref{sec:al_20k_em}. The extremely fast MAP approach is competitive with MCMC in both predictive accuracy and uncertainty quantification.

\begin{table}
    \begin{center}
    \scalebox{0.85}{\begin{tabular}{|c|c|c|c|c|c|c|c|c|c|c|c|}
    \hline
    & & \multicolumn{3}{c|}{RMSE} & \multicolumn{3}{c|}{95\% Coverage} & \multicolumn{3}{c|}{Interval Score} & Calibration Time \\
    \cline{3-12}
    & & \textit{104} & \textit{105} & \textit{106} & \textit{104} & \textit{105} & \textit{106} & \textit{104} & \textit{105} & \textit{106} & \\
    \hline
    \multirow{2}{*}{\text{Model}} & MCMC & 5.66 & 4.17 & 5.30 & 99.5\% & 100\% & 98\% & 23.14 & 22.47 & 24.06 & 52.24 minutes  \\
    \cline{2-12}
    & MAP & 4.78 & 4.50 & 6.57 & 99\% & 98.5\% & 96\% & 19.44 & 19.37 & 19.38 & 3.36 minutes \\
    \hline
    \end{tabular}}
    \caption{Calibrated prediction RMSE, median IS, and compute time for MCMC and MAP implementations of FlaGP.} 
    \label{tbl:calib_pred_obs} 
    \end{center}
    %\vspace{-8mm}
\end{table}

%For this analysis we can afford a moderate computational budget as model fitting and parameter calibration is only done once. However, an important takeaway is the relative success of the small budget MCMC and MAP calibration. For applications where new data arrives sequentially, the emulator and calibration can be updated rapidly without a notable loss in prediction accuracy or UQ.

%\vspace{-4mm}
\section{Discussion}\label{sec:discussion}
%\vspace{-4mm} 

This work introduces FlaGP, an efficient implementation of \texttt{laGP} for the emulation and calibration of computer models with functional response. FlaGP incorporates local GPs, modularization, and input scaling to enhance scalability compared to the established GP methodology described in \cite{higdon}. Unlike a fully Bayesian model that utilizes all data at each step, FlaGP offers improved efficiency while maintaining accuracy by using intelligently designed local prediction. For small datasets, the SVDGP may be preferable; however, we are frequently presented with problems for which SVDGP is computationally infeasible to implement. Through a working example, it is demonstrated that FlaGP can produce competitive predictions and may even be preferable in terms of uncertainty quantification for some data. 

Active learning for EOF-based functional response emulators is discussed in \Cref{sec:seq_d}, and a very fast approach is developed for the FlaGP emulator. It is shown how local prediction can be leveraged so that expensive computations can be recycled, leading to significantly reduced cost. For EOF-based emulators, our results indicate that using only the first (or first handful of) model component(s) may be sufficient for active learning in some cases. It is still an open question as to when this approach may fail. More generally, active learning for EOF-based functional response emulators has, to our knowledge, yet to be discussed. A careful consideration of the questions posed in \Cref{sec:seq_d} is an important area of future research. Furthermore, block sequential design may be especially relevant for economical point selection. An important extension of our active learning framework is interactive updating of the emulator with new simulator runs during Bayesian calibration such as in \citep{gu_active_learning}.
 
Importantly, it is shown that SVDGP becomes quickly infeasible in computation time with increasing $M$. The FlaGP model can be fit with nearly any computational budget by selecting an appropriate prediction neighborhood size and subset size for lengthscale estimation. This is illustrated in Section \ref{sec:al_20k_em} where model fitting and accurate prediction at 1000 points can be achieved for a large ensemble ($M=19000$) in only a few minutes.

In Section \ref{sec:al_20k_calib}, the FlaGP methodology is exercised on a large calibration problem where the SVDGP would require immense computational resources. The proposed methodology is able to constrain the input parameters using MCMC with a single observation and accurately predict an experiment using calibrated parameters in under an hour on a personal laptop. It is shown that a small computational budget is sufficient to provide accurate calibrated predictions. Furthermore, the MAP approach developed for FlaGP provides competitive calibrated predictions in under 3.5 minutes. While this method does not provide uncertainty in the calibration parameter estimates, it may be a useful tool in situations where a fast point estimate is sufficient. 

FlaGP is developed with truly large ensemble datasets in mind. For sufficiently large $M \times d_y$, computing the SVD becomes prohibitive. In this work, only data with $d_y \leq 600$ are explored; however, there exist many applications, such as spatial, image, or time series data, where $d_y$ is much larger. For the application in Section \ref{sec:Al_example}, we found that the RSVD did not lead to a less predictive emulator.

In the examples presented here, we have chosen not to incorporate strong prior information about the GP hyperparameters or the calibration parameters, but it is worth noting that incorporating stronger information may be beneficial. For example, outputs are scaled to have unit variance, so it is expected that $\sigma^2_y << 1$. If the emulator is expected to capture a significant proportion of this variability, then it may be beneficial to design a more informative prior for $\sigma^2_y$. %with nearly all of its mass below one, and a peak near $0.1$. 

%\vspace{-4mm}
\appendix
%\titleformat{\section}[block]
%  {\normalfont\Large\bfseries}{\appendixname\ \thesection}{1em}{\Large}
\section{Efficient likelihood calculation}\label{sec:eff_llh_calc}
%\vspace{-4mm}
Notice that \Cref{eqn:func_y_dist} can be rearranged as $$\bm{y}^F(\bm{x}_i^F)-\bm{B}\bm{\mu}_{w}(\bm x_i^F,\bm\theta) \sim N(\bm{0}_{d_y},\bm{B}\bm{\Sigma}_{\eta}(\bm x_i^F,\bm\theta)\bm{B}^T) + N(\bm{0}_{d_y},\sigma^2_y\bm{I}_{d_y})$$ by noting that $\bm\mu_{\eta} = \bm B\bm\mu_w$. Dropping the notation $(\bm x_i^F,\bm\theta)$ for brevity, we can similarly write \Cref{eqn:func_y_dist_biased} as $$\arraycolsep=1pt\def\arraystretch{.75}\bm{y}^F(\bm{x}_i^F)-[\bm{B};\bm{K}]\;\bigg[\begin{matrix}\bm \mu_w\\\bm \mu_v\end{matrix}\bigg] \sim \;N(\bm{0}_{d_y},\bigg[\begin{matrix}\bm{B}\bm{\Sigma}_{w}\bm{B}^T & \bm{0}\\\bm{0} & \bm{K}\bm{\Sigma}_{v}\bm{K}^T \end{matrix}\bigg]) + N(\bm{0}_{d_y},\sigma^2_y\bm{I}_{d_y}).$$ Defining $\lambda_y=\frac{1}{\sigma^2_y}$ and $\bm{r}(\bm{x}_i) = \bm{y}^F(\bm{x}_i^F)-\bm{B}\bm \mu_w$, 
\textit{Result 1.} implies that the unnormalized log likelihood in the unbiased calibration case can be written as
%\vspace{2mm}
\begin{equation}\label{eq:ll_map}
    %\vspace{2mm}
    \resizebox{.925\textwidth}{!}{% 
    $
    \begin{split}
        ll(\bm{r}(\bm{x}_i)) &\propto -\frac{1}{2}\bigg(log(|\sigma^2_{y}\bm{I}|) + log(|\lambda_y\bm{B}^T\bm{B}|) + \bm{r}(\bm{x}_i)^T\bigg[\lambda_y\bm{I} - \lambda_y\bm{B}(\lambda_y\bm{B}^T\bm{B})^{-1}\bm{B}^T\lambda_y\bigg]\bm{k}_i \bigg) + ll(\bm{\gamma}_i)\\
                &= -\frac{1}{2}\bigg((d_y-Rank(\bm{B})) log(\sigma^2_y) + log(|\bm{B}^T\bm{B}|) + \lambda_y\bm{r}(\bm{x}_i)^T(\bm{I} - \bm{B}(\bm{B}^T\bm{B})^{-1}\bm{B}^T)\bm{r}(\bm{x}_i))\bigg) + ll(\bm{\gamma}_i),% 
    \end{split}
    $
}
\end{equation}
where $\bm{\gamma}_i = \big(\bm{B}^T\bm{B}\big)^{-1}\bm{B}^T\bm{r}(\bm{x}_i) \implies \bm{\gamma}_i \sim N\big(\bm{0},\bm{\Sigma}_{w} + \sigma^2_y(\bm{B}^T\bm{B})^{-1}\big)$. For the discrepancy case, define $\bm{C}=[\bm{B};\bm{K}]$ and $\def\arraystretch{.5}\bm{r}(\bm{x}_i) = \bm{y}_i^F(\bm{x}_i^F)-\bm{C}\bigg[\begin{matrix}\bm \mu_w\\\bm \mu_v\end{matrix}\bigg]$. The log-likelihood with discrepancy can be computed using \Cref{eq:ll_map} replacing $\bm{B}$ with $\bm{C}$, and updating $\bm{r}(\bm{x}_i)$ and $\arraycolsep=1pt\def\arraystretch{.5}
\bm{\gamma}_i \sim N\big(\bm{0},\begin{bmatrix}\bm{\Sigma}_{w} & \bm{0}\\\bm{0} & \bm{\Sigma}_{v} \end{bmatrix} + \sigma^2_y(\bm{C}^T\bm{C})^{-1}\big)$ to reflect the addition of the discrepancy model. The log posterior is then $log(p(\bm{\theta}|\cdot)) \propto \sum_{i=1}^n ll(\bm{r}(\bm{x}_i)) + log(p(\bm{\theta})).$ 

\bibliography{Bibliography-MM-MC}
\end{document}

\title{Supplement: Fast Emulation and Modular Calibration for Simulators with Functional Response}
%\author{Grant Hutchings \& Derek Bingham}

\renewcommand{\thesection}{SM\arabic{section}}
\renewcommand{\thesubsection}{SM\arabic{section}.\arabic{subsection}}
\renewcommand{\thefigure}{SM\arabic{figure}}
\renewcommand{\thetable}{SM\arabic{table}}
\renewcommand{\theequation}{SM\arabic{equation}}

\author[cor1,1,2]{Grant Hutchings}
\author[2]{Derek Bingham}
\author[1]{Earl Lawrence}

%% Author affiliation

\address[1]{Statistical Sciences, Los Alamos National Laboratory, NM, United States}
\address[2]{Department of Statistics, Simon Fraser University, BC, Canada}

\cortext[cor1]{Corresponding author: Email: grant.hutchings@lanl.gov (Grant Hutchings), Phone: +1 505-667-2200, Present address: P.O. Box 1663 Los Alamos, NM 87545.}

\maketitle

\section{FlaGP Algorithm in detail}

A detailed algorithm the implementation of the methods presented in this manuscript is given here, which are implemented in the \texttt{R} package \href{https://github.com/granthutchings/FlaGP/tree/main}{\texttt{FlaGP}}. Algorithm 1. in the manuscript provides an outline for fitting the emulator and making predictions. Algorithms 2. \& 3. define the procedures for using predictions from the emulator in a modular calibration framework. The purpose of these supplemental algorithms is to provide more detail that should be sufficient to implement the methods from scratch.

\begin{breakablealgorithm}
\caption{Fast Emulator fit and Prediction}\label{flagp_alg}
\begin{algorithmic}[1]\vspace{2mm}
\Algphase{Fitting the FlaGP Emulator}
\Statex \textbf{Precomputing 1: Data Setup}
\vspace{2mm}
%\Require $\bm{X}$, $\bm{Z}$
\State Scale input matrix $\bm{X} \in \mathbb{R}^{M \times d_x}$ to unit Hypercube $[0,1]^{d_x}$.
\State Center output Matrix $\bm{Z} \in \mathbb{R}^{d_y \times M}$ by subtracting the mean at each functional index. Optionally, scale $\bm{Z}$ to have unit variance.
\State Compute singular value decomposition $\bm{Z}=\bm{UDV^T}$
\State Select $p_{\eta}$, the number of basis components to keep for modeling, by choosing a minimum desired proportion of variability in $\bm{Z}$ to be explained by the decomposition (commonly $.95$). The proportion of variance explained by each singular vector is $\bm{d}^2/\sum \bm{d}^2$ where $\bm{d}$ is the vector of singular values, the diagonal elements of $\bm{D}$.
\State Define the low rank basis $\bm{Z} \approx \bm{B}\bm{W}(\bm{X})$ where $\bm{B}$ is the first $p_{\eta}$ columns of $\bm{U}\bm{D}/\sqrt{M}$ and $\bm{W}(\bm{X})$ is the first $p_{\eta}$ rows of $\bm{V}^T\sqrt{M}$.
%\EndProcedure%
%\Procedure{Estimate GP correlation parameters}{}
\vspace{2mm}
\Statex \textbf{Precomputing 2: Estimate emulator hyperparameters}
\vspace{2mm}
\State \textit{Option 1}: For each of the $p_{\eta}$ rows in $\bm{W}(\bm{X})$, estimate lengthscales (and optionally a nugget $g$) using the \texttt{laGP} function \texttt{blhs.loop} with $d_{est}$ divisions on each coordinate in $\bm{X}$ and $r_{est}$ bootstrap samples. This function uses Bootstrapped Block Latin hypercube sampling to define $r_{est}$ data subsets, then estimates correlation lengths for each subset using an Empirical Bayes MAP estimation procedure. The median over bootstrap subsets are used as estimates $\{\bm{\hat{l}}\}_{p_{\eta}}$.
\State \textit{Option 2}: Sample $r_{est}$ bootstrap subsets of size $m_{est}$ using stratified or simple random sampling. Use the \texttt{laGP} functions \texttt{newGPsep} and \texttt{mleGPsep} to estimate correlation lengths using the same Empirical Bayes procedure for each bootstrap subset. Take the median over subsets to estimates $\{\bm{\hat{l}}\}_{p_{\eta}}$.\newline

\noindent The computation time of the estimation procedure is dependent on $d_{est}$ or $m_{est}$ and $r_{est}$. In practice, we recommend setting $d_{est}$ small and $m_{est},r_{est}$ large if computational budget allows.
\vspace{2mm}
\Statex \textbf{Precomputing 3: Stretch and Compress inputs} 
\vspace{2mm}
\State Given estimates $\bm{\hat{l}}_j;\;j=1,\ldots,p_{\eta}$, defined stretched and compressed inputs \[\bm{X}^{sc}_j=\bm{X}/\sqrt{\hat{\bm{l}}_j},\]where division is column-wise, that is, the $k$th column of $\bm{X}^{sc}_j$ is computed by dividing the $k$th column of $\bm{X}$ by the square root of the $k$th lengthscale estimate for the $j$'th basis component $\hat{l}_{kj}$. That is, \[{x}_{kj}^{sc} = {x}_{kj}/\sqrt{\hat{l}_{kj}}.\]
\Algphase{Predictions from fitted Emulator at $\bm x^*$}
\State Compute $p_{\eta}$ scaled prediction inputs \[\bm{x}^{*sc}_j=\bm{x}^*/\sqrt{\hat{\bm{l}}_j}.\]
\State Build nearest neighbor training datasets: Define training data for the $j$th basis component model as the $m$ nearest neighbor inputs to $\bm{x}^{*sc}_j$ in $\bm{X}^{sc}_j$ along with their associated basis weights from $\bm{W}(\bm{X})$. Nearest neighbors are computed using a fast K-d tree algorithm implemented in the \texttt{R} function \texttt{FNN::get.knnx()}.
\State Compute predictive distribution: Training data defines predictive distributions (Equation 3. in the manuscript) for for $w_j(\bm{x}^{*sc}_j)$.
\State Combine predictions from $p_{\eta}$ basis models: To compute predictions in function space, propagate predictive means $\mu_j(\bm x^{*sc})$, or samples $\tilde{w}_j(\bm x^{*sc}_j)$, through the basis vectors to get \[\tilde{\bm{z}}(\bm{x})=\sum_{i=1}^{p_{\eta}} \bm{b}_j\tilde{w}_j(\bm{x}^{*sc}).\]
\end{algorithmic}
\end{breakablealgorithm}
\newpage
\begin{breakablealgorithm}
\caption{Modular Calibration with FlaGP Emulator}\label{flagp_alg_2}
\begin{algorithmic}[1]\vspace{2mm}
\Require \textit{Fitted Emulator}. 

\noindent In this case, the two types of inputs $\bm{X}$, $\bm{T}$, require $d_x+d_t$ lengthscale parameters to be estimated for each of the $p_{\eta}$ models. Denote two sets of lengthscale parameters as $\{\bm{\hat{l}}_x\}_{p_{\eta}}$, $\{\bm{\hat{l}}_t\}_{p_{\eta}}$.
\vspace{2mm}
\Statex \textbf{Sample candidate parameters $\bm t$, $\sigma^2_y$}
\vspace{2mm}
\State Given an MCMC algorithm, propose a set of candidate parameters $\bm t$ and $\sigma^2_y$ given some initial state $\bm t_{\text{init}}$, $\sigma^2_{y,\text{init}}$.
\vspace{2mm}
\Statex \textbf{Predict Field Observations at} $\bm{\theta} = \bm{t}$
\vspace{2mm}
\State Stretch and compress inputs: Compute stretched and compressed inputs $\{\bm{X}_F^{sc}\}_{p_{\eta}}=\bm X_F/\sqrt{\{\bm{\hat{l}}_x\}_{p_{\eta}}}$ and $\{\bm{t}^{sc}\}_{p_{\eta}} = \bm t/\sqrt{\{\bm{\hat{l}}_t\}_{p_{\eta}}}$
\State Emulator Prediction: Acquire samples from the predictive distributions of $w_j(\bm{x}_{i,j,F}^{sc},\bm{t}^{sc}_j)\;j=1,\ldots,p_{\eta},\;i=1,\ldots,n$ using the procedure in Algorithm \ref{flagp_alg}. Letting $\tilde{\bm W}$ denote the matrix of predictive samples, form the matrix of residuals $\bm{R} = \bm{Y} - \tilde{\bm{W}}\bm{B}$.

\State Discrepancy Model: With user-defined discrepancy basis matrix $\bm{D}$, compute projected residual matrix $\bm{V} = (\bm{D}^T\bm{D})^{-1}\bm{D}^T\bm{R}$. Fit independent full Gaussian Process models to the rows of $\bm{V}$, each with input matrix $\bm{X}_F$ using \texttt{laGP} functions \texttt{newGPsep} and \texttt{mleGPsep}. Stretched and compressed inputs are not used for the discrepancy model. Compute the  the predictive mean and variance of the of the discrepancy model at each of $n$ the observed inputs $\bm{X}_F$. Denote samples from the predictive distribution as $\tilde{\bm{v}}_i$.
\vspace{2mm}
\Statex \textbf{Compute log-posterior at $\bm{\theta}=\bm{t}$}
\vspace{2mm}
\State Compute log-posterior: According to the model \[\begin{aligned}
    \bm{y}(\bm{x}_i^F)\;|\bm \eta(\bm x_i^F,\bm\theta),\bm{\delta}(\bm{x}_i^F),\bm{\theta},\sigma^2_y \sim N(&\bm\mu_{\bm{\eta}}(\bm{x}_i^F,\bm{\theta}) + \bm\mu_{\bm\delta}(\bm x_i^F),\\&\bm B^T\bm \Sigma_{\bm \eta}(\bm x_i,\bm\theta)\bm B +\\& \bm K^T\bm \Sigma_{\bm \delta}(\bm x_i)\bm K + \sigma^2_y\bm{I}_{d_y});\;i=1,\ldots,n,
    \end{aligned}\] compute posterior \[p(\bm{\theta},\sigma^2_y|\cdot) \propto \prod_{i=1}^n p(\bm{y}(\bm{x}_i)|\bm{\eta}(\bm x_i,\bm\theta),\bm{\delta}(\bm x_i),\bm{\theta},\sigma^2_y) \times p(\bm{\theta}) \times p(\sigma^2_y)\] where $\bm r_i = \bm{y}_i(\bm{x}^F_i) - \bm \eta(\bm x_i^F, \bm \theta) - \bm \delta(\bm x_i^F) = \bm{y}_i(\bm{x}^F_i) - \tilde{\bm w}_i\bm B - \tilde{\bm v}_i\bm D$.

\State Repeat steps 1-5 in an MCMC loop to sample from $p(\bm{\theta},\sigma^2_y|\cdot)$.
\end{algorithmic}
\end{breakablealgorithm}

\section{Replacing points in a prediction neighborhood}\label{sec:rep_nbhd}
	
We give a simple example illustrating that when replacing a point in a nearest neighbor prediction set the prediction accuracy is not guaranteed to be reduced, even though the added point is closer to the prediction point. This phenomenon can be observed even outside of two important scenarios where it might be expected, that is when
\begin{itemize}
    \item The point replacement results in the prediction point being an extrapolation
    \item The new point is not closer in scaled distance to the prediction point than the point it replaced. This could occur if the nearest neighbors are not selected with respect to the distance metric under the posterior GP's covariance matrix.
\end{itemize}

In the first scenario, we would not be surprised at all to see prediction variance increasing when the point becomes an extrapolation. The second scenario is more complex, but a simple example may suffice to describe the setting. Suppose we have a scalar response function $y(x_1,x_2)=x_1$. Fitting a GP to this function with a separable covariance structure, we would expect that the lengthscale parameters would indicate that correlation in the response is high w.r.t changes in $x_1$, but very low (0) w.r.t. changes in $x_2$. Suppose that we wished to make a prediction at say, $\bm x_p=(.5,.5)$ with a neighborhood size of 2. Let the current neighborhood contain the points $\bm x^1=(.4,.4)$ and $\bm x^2=(.6,1)$. If we consider adding the point $\bm x^3=(.7,.4)$ to the neighborhood, which would kick $x^2$ out, it would not be at all surprising if prediction error increased because we have sacrificed closeness of training data in the direction that is important for training data that is closer in absolute distance but will provide less relevant information about the function.

\begin{figure}
    \centering
    \includegraphics[width=\linewidth]{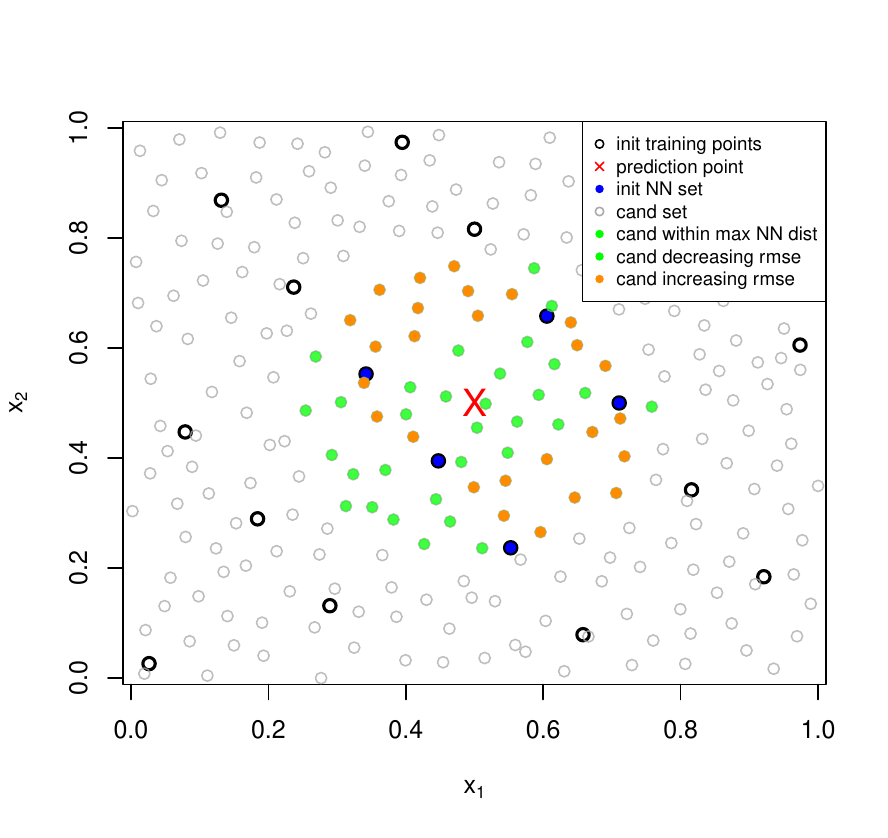}
    \caption{Caption}
    \label{fig:bad_nbhd}
\end{figure}

We were somewhat surprised to find that the phenomenon of increasing prediction error can happen when the replaced point does not make for an extrapolation, and, the kernel function is isotropic. We illustrate this with a simple example where data is generated from the function
$$y(x_1,x_2) = x_1+x_2,$$
so both inputs are equally important to the response both in absolute terms, and according to the GP kernel function 
$$K(\bm x,\bm x') = \sum_{j=1}^2 \sqrt{(x_j-x'_j)^2}.$$
All the relevant information for this example is contained in \Cref{fig:bad_nbhd}. The red \text{X} indicates the point where a prediction will be made, and the thick black circles are the entire training set with which to select nearest neighbors from. Some of these points, 5 to be exact, are colored in blue, indicating that they are the 5 nearest neighbors according to the isotropic correlation function given above. Now suppose that we have a large set of candidate points which might be added to the training data. Under this correlation function, the points which could potentially replace a point in the current nearest neighbor set are colored as green and orange. These are the points that are closer to the prediction location than the furthest blue point. The point colored orange are those that actually increase prediction error at the red point, while the green points are the ones which reduce prediction error. It is paradoxical that so many of these points would increase prediction error given that they do not cause an extrapolation at the prediction point in either direction. Any of the candidate points that are added to the neighborhood remove the blue point at $(0.55,0.24)$, but there does not appear to be a clear pattern as to which points increase or decrease prediction error with respect to the removed point.

\section{Computational speedups from input scaling}

A simple simulation study is presented to demonstrate the computation gains that can be expected from input scaling with nearest neighbors compared to a default implementation of laGP. In the manuscript, two important benefits of input scaling are discussed; First, by scaling inputs, nearest neighbor subsets can be dramatically improved under an anisotropic correlation function. This allows us to forgo computing optimal prediction neighborhoods as is done in laGP. By default laGP uses the ALC criterion to select neighbors, which we will explore in this study. Second, scaling inputs allows for the lengthscales to be estimated only once, rather than at every prediction. In sequential prediction tasks (like computer model calibration) this can be a very important reduction in computational cost as we will see. The benefit of non-stationarity in predictions is lost, but the simulation study indicates that speedups due to input scaling could be as large as 2 orders of magnitude (100x) when using the default ALC criterion with parameter estimation compared to nearest neighbors with pre-estimated lengthscales. For our applications, a small reduction in predictive accuracy is acceptable given these computational tradeoffs.

\subsection{Neighborhood selection with laGP}

\begin{figure*}
    \centering
    \begin{subfigure}[b]{0.45\textwidth}
        \centering
        \includegraphics[width=\textwidth]{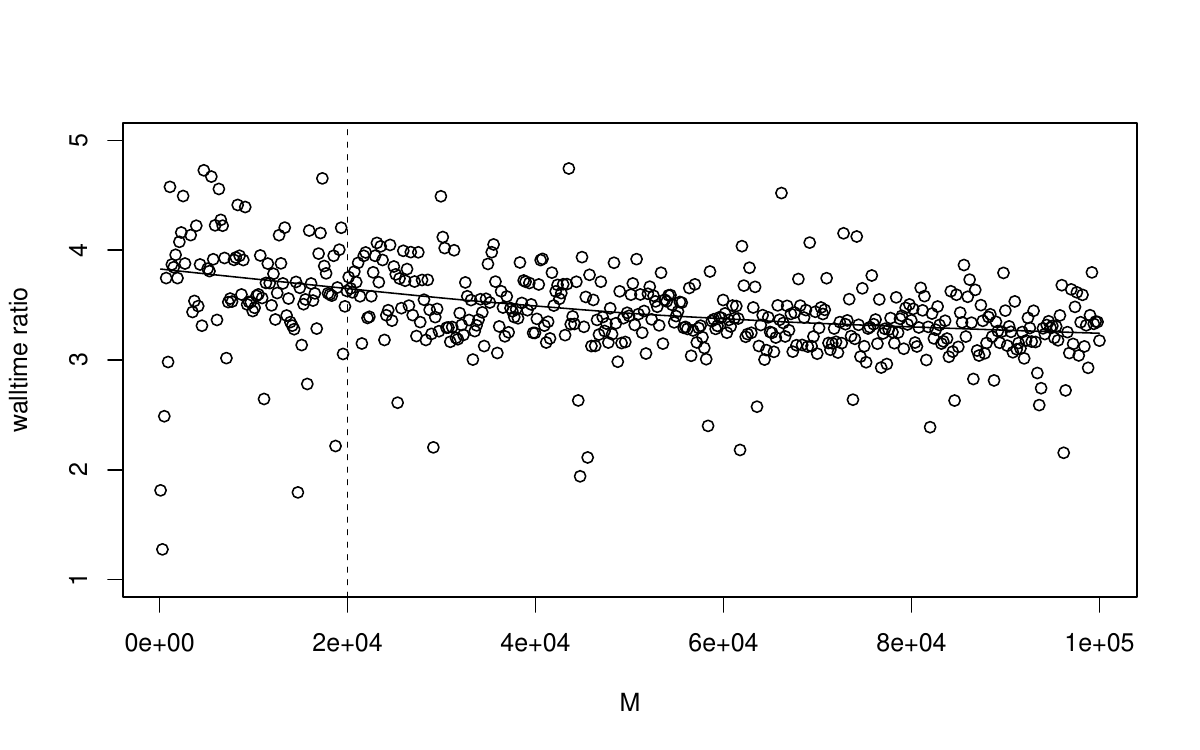}
        \caption{\small Ratio as a function of ensemble size ($M$).}
        \label{fig:}
    \end{subfigure}
    \begin{subfigure}[b]{0.45\textwidth}  
        \centering 
        \includegraphics[width=\textwidth]{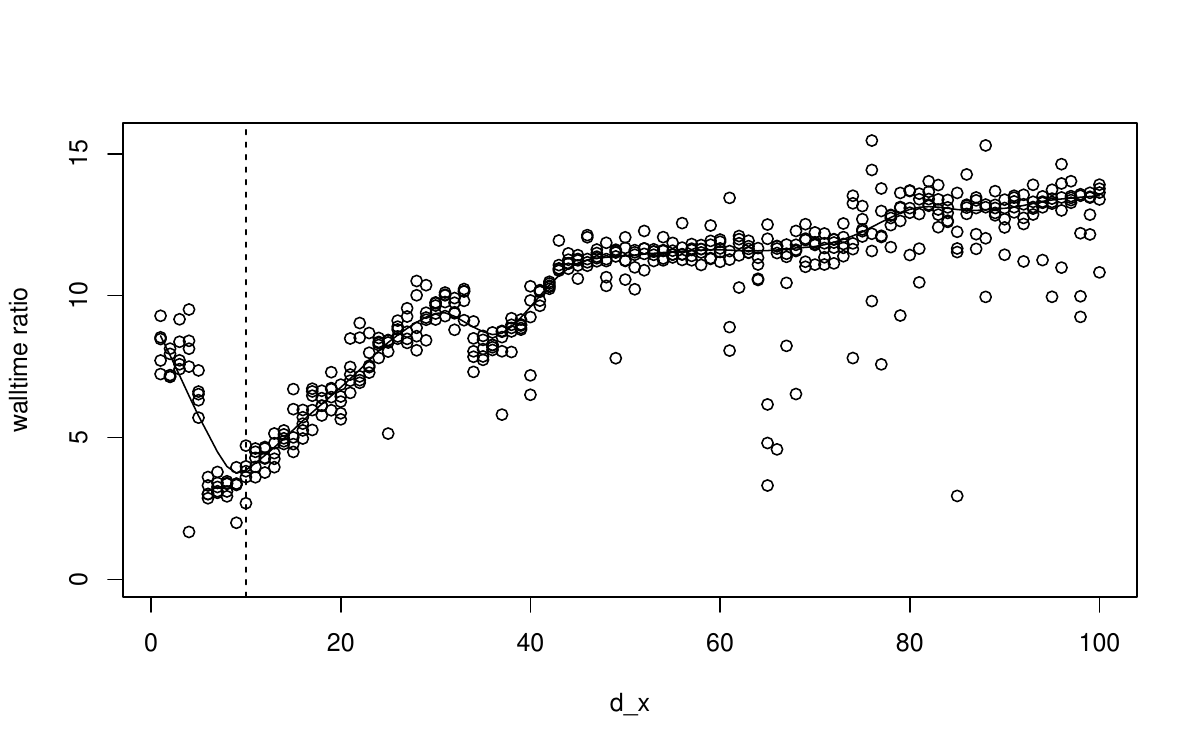}
        \caption{\small Ratio as a function of input dimension ($d_x$).}   
        \label{fig:}
    \end{subfigure}
    \caption{\small Comparison of wall time for ALC neighborhood selection compared to simple nearest neighbors. For fixed $d_x$, ratio decreases slightly with increasing $M$, and for fixed $M$ ratio increases with increasing $d_x$ after about $d_x=5$. Vertical black lines indicate the approximate size of the application dataset.}
    \label{fig:alc_nn}
\end{figure*}

The goal of this study is to quantify the relative cost of the ALC criterion over a range of data set sizes. First, we fix the dimension of the input space at $d_x=10$ and generate 500 random datasets with $100 \leq M \leq 100000$. Input and output data $\bm{X},\; \bm{y}$ are generated uniformly on the unit hypercube. Predictions are made sequentially at 10 locations using both the ALC and NN criteria. Input scaling is not used, so correlation parameters must be estimated for each prediction. Additionally, 500 datasets are generated with fixed $M=20,000$ and $1 \leq d_x \leq 100$ with 5 repeated datasets generated for each value of $d_x$. We see that ALC is between 3 and 5 times slower than NN over a large range of $M$ at $d_x=10$ and increases for larger $d_x$. There is significantly more variability in the ratio as we change $d_x$ while keeping $M$ fixed, with ratios ranging from 3 to 15. Vertical black lines in Figure \ref{fig:alc_nn} indicate datasets with size representative of the application data set ($M=20000,d_x=10$).

\subsection{Parameter Estimation with laGP}

\begin{figure*}
    \centering
    \begin{subfigure}[b]{0.45\textwidth}
        \centering
        \includegraphics[width=\textwidth]{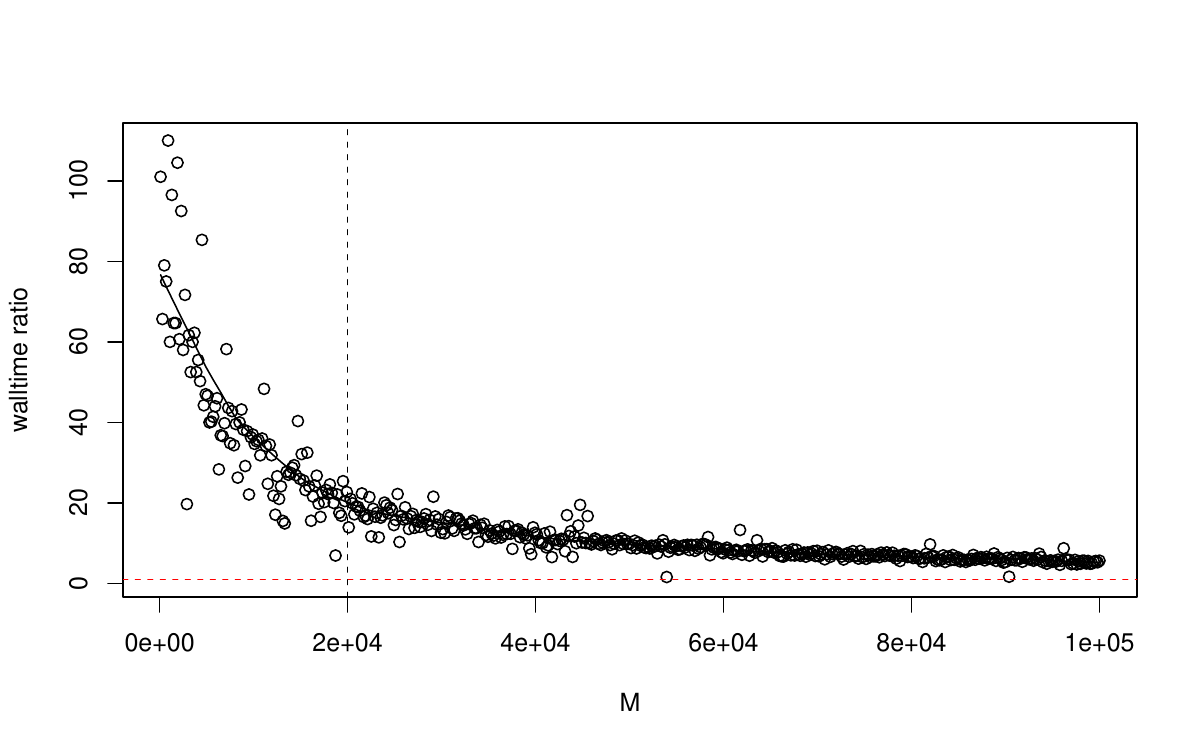}
        \caption{\small Ratio as a function of ensemble size ($M$).}
        \label{fig:}
    \end{subfigure}
    \begin{subfigure}[b]{0.45\textwidth}  
        \centering 
        \includegraphics[width=\textwidth]{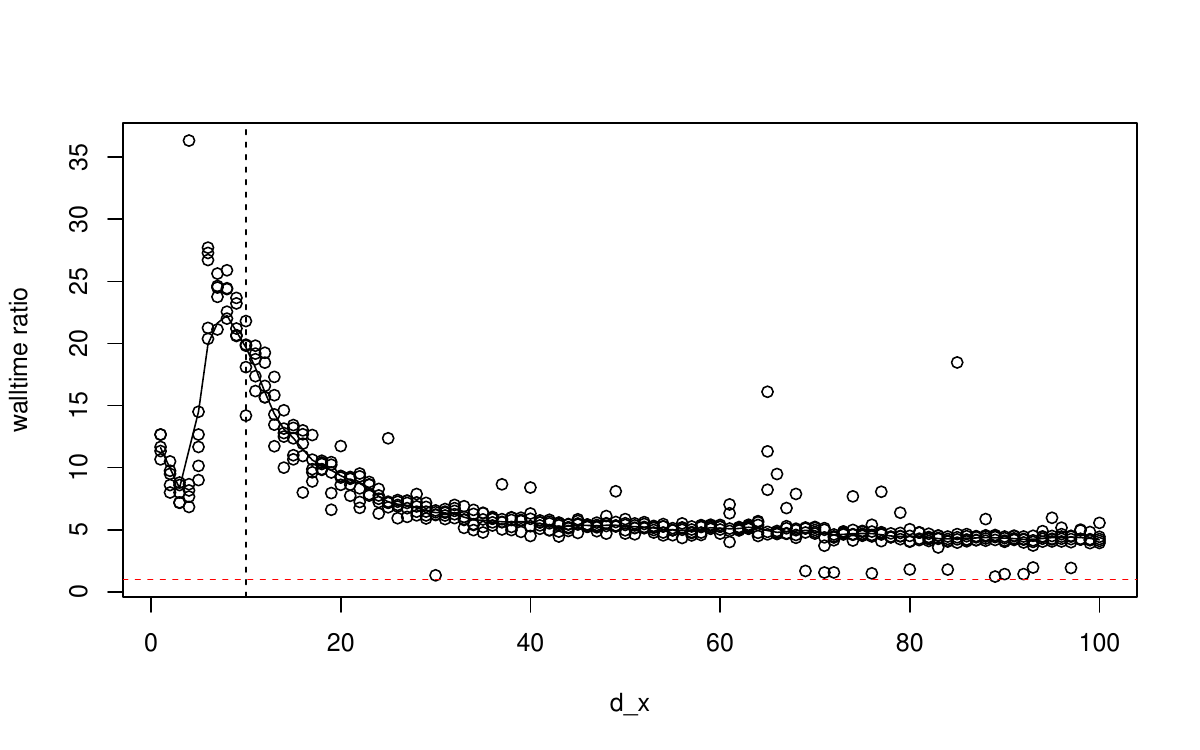}
        \caption{\small Ratio as a function of input dimension ($d_x$).}   
        \label{fig:}
    \end{subfigure}
    \caption{\small Comparison of computation time for nearest neighbor selection with and without repeated parameter estimation. }
    \label{fig:nn_est_par}
\end{figure*}

The cost of repeated parameter estimation during prediction, even for small $m$, can add up to be significant over thousands of MCMC iterations. A similar simulation study is performed comparing prediction time for the NN criteria with local parameter estimation, and the NN criteria assuming that global lengthscales have been pre-estimated, and the results are shown in Figure \ref{fig:nn_est_par}. As both $M$ and $d_x$ increase, the ratio appears to be converging to 1 (horizontal red line). This is because the computation time associated with finding nearest neighbors dominates that of parameter estimation for large data. For $M=20,000$ and $d_x=10$ (vertical black lines), which approximately represents the application dataset, speedups are still very significant at (about $20\times$). Assuming a $5\times$ speedup from NN vs ALC, and a $20\times$ speedup from pre-estimated lengthscales, we have approximates a 2 order of magnitude speed increase. Obviously, the time required to pre-estimate the lengthscales should not be ignored, but even for the large application dataset considered the cost of pre-estimation is very small compared to the cost of sequential prediction in during Bayesian calibration.

